\documentclass[aps, prb, twocolumn, showpacs, superscriptaddress, floatfix]{revtex4-1}

\usepackage[utf8]{inputenc}
\usepackage{amsmath, amssymb}
\usepackage{commath}
\usepackage{graphicx}
\usepackage{hyperref}
\usepackage{color}

\usepackage[makeroom]{cancel}
\usepackage{soul}
\usepackage{stmaryrd}

\def\be{\begin{equation}}
\def\ee{\end{equation}}

\newcommand{\hnd}[1]{ \hat{\mathrm{{\textbf{#1}}}} }

\begin{document}

\title{Reconstructing non-equilibrium regimes of quantum many-body systems from
the analytical structure of perturbative expansions}
\author{Corentin Bertrand}
\affiliation{Univ. Grenoble Alpes, CEA, INAC-PHELIQS, GT  F-38000 Grenoble, France}
\author{Serge Florens}
\affiliation{Univ. Grenoble Alpes, CNRS, Institut N\'eel, F-38000 Grenoble, France}
\author{Olivier Parcollet}
\affiliation{Center for Computational Quantum Physics, Flatiron Institute, 162 5th Avenue, New York, NY 10010, USA}
\affiliation{Institut de Physique Th\'eorique (IPhT), CEA, CNRS, UMR 3681, 91191 Gif-sur-Yvette, France}
\author{Xavier Waintal}
\affiliation{Univ. Grenoble Alpes, CEA, INAC-PHELIQS, GT F-38000 Grenoble, France}

\begin{abstract}
We propose a systematic approach to the non-equilibrium dynamics of strongly
interacting many-body quantum systems, building upon the standard perturbative expansion 
in the Coulomb interaction. High order series are derived from the Keldysh version of 
determinantal diagrammatic Quantum Monte Carlo, and the reconstruction beyond the weak 
coupling regime of physical quantities is obtained by considering them as analytic 
functions of a complex-valued interaction $U$. Our advances rely on two crucial 
ingredients: i) a conformal change of variable, based on the approximate 
location of the singularities of these functions in the complex $U$-plane;
ii) a Bayesian inference technique, that takes into account additional
known non-perturbative relations, in order to control the amplification of noise 
occurring at large $U$. This general methodology is applied to the strongly
correlated Anderson quantum impurity model, and is thoroughly tested both in- and 
out-of-equilibrium. In the situation of a finite voltage bias, our method is able 
to extend previous studies, by bridging with the regime of unitary conductance, and
by dealing with energy offsets from particle-hole symmetry. We also confirm the 
existence of a voltage splitting of the impurity density of states, and find that 
it is tied to a non-trivial behavior of the non-equilibrium distribution function.
Beyond impurity problems, our approach could be directly applied to Hubbard-like 
models, as well as other types of expansions.
\end{abstract}
 
\maketitle

\section{Introduction}

The study of the out-of-equilibrium regime of strongly correlated many-body quantum problems is a major challenge in theoretical condensed matter physics.
Its interest has grown rapidly in the past few years with new
experiments, \textit{e.g.} the ability to control 
light-matter interaction on ultra-fast time scale\cite{Foerst2011}, light-induced
superconductivity \cite{Fausti2011, Nicoletti2014, Casandruc2015, Nicoletti2016, Nicoletti2018} or metal-insulator transition driven by
electric field \cite{Nakamura2013}, proposed \textit{e.g.} to build artificial neurons\cite{Valle2018}.  
These experiments raise the question whether the combination of strong correlation effects and out of equilibrium
regimes could lead to genuinely new physics and phases of matter that do not have an
equilibrium counterpart. 
Quantum nanoelectronics also provide many examples of such systems.
A classic example is the spin-1/2 Kondo effect occurring in a quantum dot, but recent experiments have 
also managed to study in great detail underscreened~\cite{Roch2009,Parks2010} and
overscreened~\cite{Iftikhar2015,Iftikhar2018} (multi-channel) Kondo effects, characterized by non-Fermi liquid 
fixed points.
Other notable examples of new quantum states induced by interactions are
Luttinger liquids\cite{Giamarchi2004} that take place 
at edges in the fractional quantum Hall regime, or the ``0.7
anomaly"\cite{Thomas1996, Thomas1998, Micolich2011} occurring in a simple
quantum point contact geometry.
Last, solid state based quantum computers such as spin qubits devices are nothing but out-of-equilibrium quantum 
many-body systems (few sites Hubbard like models, possibly connected to electrodes) that bring new questions into 
the scope of correlated systems\cite{Preskill2018}.   

It is worth noting that even the simplest of these out-of-equilibrium problems,
the single impurity Anderson model, is still the subject of active
research\cite{Reininghaus2014, Schwarz2018}. 
Early approaches used a range
of approximate techniques including 4th order perturbation
theory\cite{Fujii2003}, equation of motion techniques\cite{VanRoermund2010} and
the Non Crossing Approximation (NCA)\cite{Wingreen1994}.  
State of the art techniques include the time-dependent Numerical Renormalization Group (NRG) and the
density matrix renormalization group (DMRG)\cite{White1992, White1993, Schollwoeck2005, Anders2005,
Heidrich-Meisner2009, Schwarz2018, Eckel2010}. 
Early attempts of real time quantum Monte Carlo \cite{Muehlbacher2008, Werner2009, Werner2010, Schiro2009, Schiro2010}
have experienced an exponential sign problem at long time and large interaction.
Within Monte-Carlo methods, two main routes are currently explored to resolve this issue:
the inchworm algorithm \cite{Cohen2014a, Cohen2014b, Cohen2015, Chen2017a, Chen2017b}
and the Schwinger-Keldysh diagrammatic Quantum Monte Carlo \cite{Profumo2015} (QMC).
The later, which we use in this paper, reaches the infinite time steady state limit 
and has a complexity which does not grow with time.
The development of controlled computational methods is critical for the development of the theory
in this field. Beyond its direct application to impurities and quantum dot
physics, the Anderson model is of direct interest for quantum embedding methods
such as Dynamical Mean Field Theory \cite{Georges1996, Kotliar2006, Aoki2014} (DMFT)
which reduce bulk lattice problem to the solution of a self-consistent quantum
impurity model.

A straightforward approach to study out-of-equilibrium many-body quantum problem
is to compute the systematic perturbative expansion of some physical quantity
$F$ in power of the electron-electron interaction $U$: $F(U) \equiv
\sum_{n=0}^\infty F_n U^n$. In practice, $F$ may depend on time (or frequency)
as well as voltage-bias, temperature, \textit{etc}.
The coefficients $F_n$ are given by the out-of-equilibrium Schwinger-Keldysh version of the Feynman diagrams\cite{Rammer2007}.
Such a perturbative expansion is a central tool in quantum mechanics and quantum field theory.  
In weak coupling theories, a few orders are sufficient to explain many physical phenomena, even
quantitatively, as \textit{e.g.} in Quantum Electrodynamics (QED). 
However, at intermediate or strong coupling, this approach faces two main challenges: 
{\it (i)} the computation of the coefficients for $n$ large enough and 
{\it (ii)} the reconstruction of the physical quantities as a function of $U$ from a finite number of coefficients.

Using the standard Wick theorem, an explicit expression of $F_n$ to order $n$
can be written as $n-$dimensional integrals.
While the computation of $F_n$ can hardly been achieved analytically beyond a few orders, 
Quantum Monte-Carlo (QMC) algorithms known as ``diagrammatic
Monte-Carlo"\cite{Prokofev1998, Mishchenko2000, VanHoucke2008, Prokofev2007,
Prokofev2008, Gull2010, Kozik2010, Pollet2012, VanHoucke2012, Kulagin2013,
Kulagin2013a, Gukelberger2014, Deng2015, Huang2016, Rossi2018a, VanHoucke2019} are able to 
compute a finite number of these coefficients $F_n$ for a general class of
quantum many-body problems, in practice up to $8$ or $15$ depending on the model and the physical quantity.
The first generation of these algorithms
explicitly sampled the Feynman diagrams one by one with a complex Markov chain,
moving from one diagram to another.  A second generation of algorithms handles the diagrams
collectively using combinations of determinants to cancel disconnected diagrams
in physical quantities.
This was achieved in the real time Schwinger-Keldysh formalism\cite{Profumo2015},
and in the imaginary time Matsubara formalism\cite{Rossi2017b, Moutenet2018, Simkovic2017, Rossi2018b}.

The resummation of the series is a non-trivial mathematical task outside of the weak
coupling regime, even with a perfect knowledge of the coefficients $F_n$.  
The issue comes from the finite radius of convergence of the series.  When $U$ is
larger than this radius, the truncated series to the first $N$-th terms does not converge with $N$
and some resummation technique must be used to compute $F(U)$.
Moreover, there are two additional difficulties associated with numerical methods: 
{\it i)} only a finite number of coefficients $F_n$ can be computed since
the computation cost is exponential in $n$ and {\it ii)} the $F_n$ are
only known with a finite precision, typically of a few digits in QMC.

\begin{figure}[t]
    \centering
    \includegraphics[width=8cm]{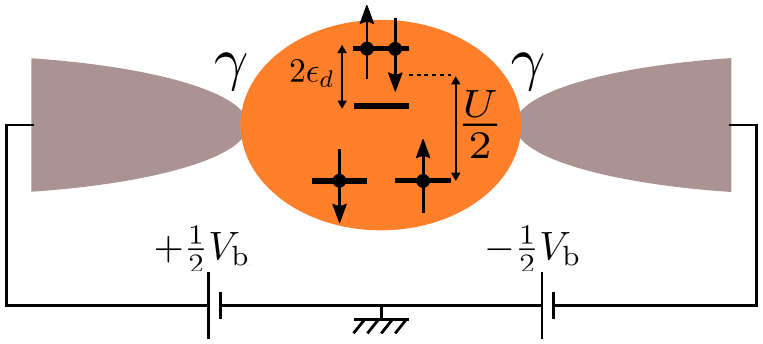}
    \includegraphics[width=8cm]{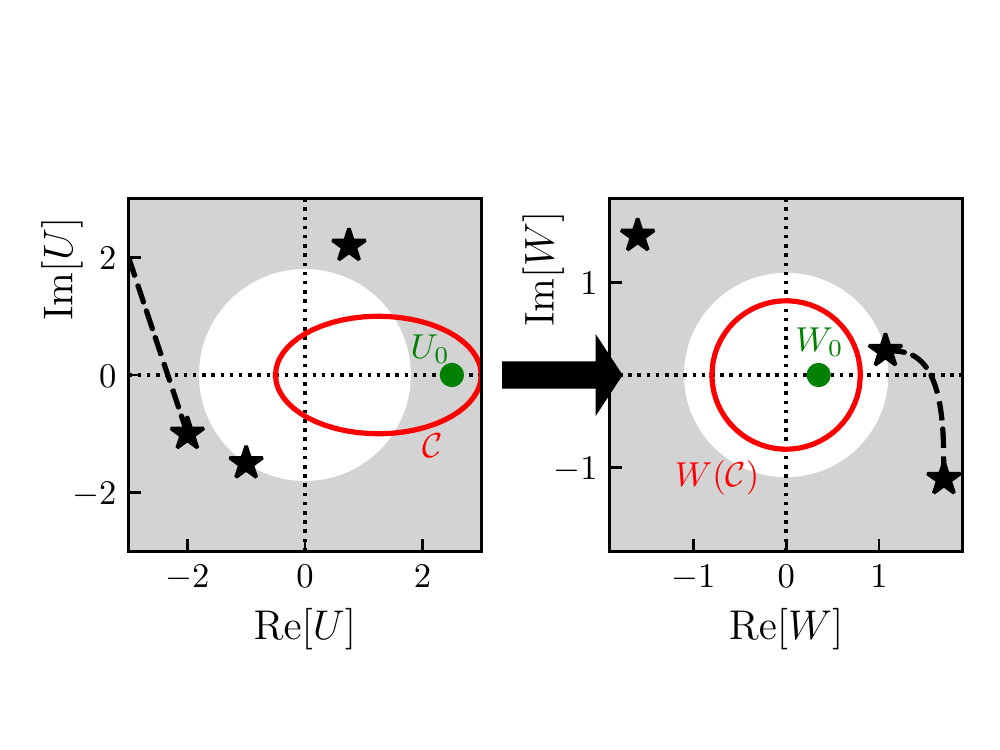}
    \caption{
    \label{fig:intro}
    Upper panel: the Anderson quantum impurity model describing
    a single level quantum dot. The level with energy $\epsilon_d$ is subject to 
    a finite Coulomb interaction $U$, and is hybridized with a tunnel coupling $\gamma$ 
    to two leads that are biased with voltage $V_b$.
    Lower left panel: 
    illustration of the general computation scheme developed in this work. 
    A physical quantity $F$ (\textit{e.g.} the current through the dot) presents
    singularities in the $U$ complex plane, such as poles (stars) or branch cuts
    (dashed line), hampering proper convergence of perturbative approaches for
    values of $U$ outside the convergence disk (grey area). After defining a broad
    singularity-free contour $\cal C$ (red line) that encircles both $U=0$ and a 
    targeted $U_0$ value, a conformal map $U\to W(U)$ is defined in order to
    bring $W_0=W(U_0)$ inside the convergence disk of $F[U(W)]$ (lower right
    panel). Resummation techniques can then be applied in a controlled way.
}
\end{figure}

In this paper, we approach this problem from the angle of complex analysis.
Indeed, the divergence of the series originates from the singularity structure 
of the function $F(U)$ in the complex plane $U$ (lower left panel in
Fig.~\ref{fig:intro}). We discuss how to locate the singularities closest to 0,
and how to construct an analytic change of variable to resum the series beyond weak coupling (lower 
right panel in Fig.~\ref{fig:intro}). We also introduce a Bayesian technique to take into account the 
amplification of the Monte-Carlo noise in the resummation process using some simple non-perturbative 
additional information on the model.

While our approach is quite general, we will focus here on the non-equilibrium Anderson quantum impurity model
in the quantum dot configuration (upper panel in Fig.~\ref{fig:intro}).
Our starting point is an expansion of the Green's function in power of the
Hubbard interaction $U$, using an extension of the algorithm of Ref.\cite{Profumo2015}.
The algorithm is discussed in details in a companion paper \cite{Bertrand2019a}, 
its implementation is based on the TRIQS library\cite{Parcollet2015}.
This algorithm provides a numerically
exact computation of the perturbative series of
physical quantities in power of the interaction $U$, at a cost which is {\it uniform in time} but exponential with the expansion order.
Hence it allows to compute in a transient regime as well as directly in a long time steady state, a regime in which most competing methods have severe limitations. 

This paper is organized as follows.
Section~\ref{sec:model} introduces our notations for the single impurity
Anderson model. Section~\ref{sec:resummation} develops the resummation technique
and illustrates it on the Kondo temperature. Section~\ref{sec:bench} performs a
benchmark of the method against NRG for the equilibrium dynamics. Section~\ref{sec:results} 
presents new results in the non-equilibrium regime, including the voltage-split spectral 
function, extended-range current-voltage characteristics, and a non-trivial dot distribution 
function.
Section~\ref{sec:conclusion} concludes this article and presents perspectives for
our conformal approach to the perturbative expansions of strongly interacting
quantum systems.

\section{The Anderson impurity model}
\label{sec:model}

In this paper, we focus on the single impurity Anderson model both at
and out-of equilibrium.  While originally formulated to describe the effect of
magnetic impurities in metals, this model is widely used in theoretical
condensed matter, both as a simple model for quantum dots in mesoscopic
physics and as a building block of ``quantum embedding"  approximations like
DMFT and its generalizations.  At the core of the
Anderson model lies Kondo physics. The repulsive interaction on the quantum dot leads
to an effective antiferromagnetic interaction between the
electronic reservoirs and the spin of the (unique) electron trapped in the
quantum dot in the local moment regime. This interaction leads to the formation of the Kondo resonance, a
thin peak in the local density of state pinned at the Fermi
energy\cite{Hewson1993}. 
The Anderson impurity Hamiltonian reads:
\begin{eqnarray}
\hnd{H}
&=& \sum_{i=-\infty}^{+\infty}\sum_\sigma \gamma_i
\hnd{c}^\dagger_{i,\sigma}\hnd{c}_{i+1,\sigma}+h.c.
+\epsilon_d (\hnd{n}_\uparrow+\hnd{n}_\downarrow)
\nonumber \\
\label{eq:model}
&& +U\theta(t)\left(\hnd{n}_\uparrow-\frac{1}{2}\right)\left(\hnd{n}_\downarrow-\frac{1}{2}\right).
\end{eqnarray}
It connects an impurity on site $0$ to two semi-infinite electrodes $i<0$ and
$i>0$.  The model corresponds to a single level artificial atom as sketched in the upper panel of 
Fig.~\ref{fig:intro}. Here $\epsilon_d$ is the on-site energy
of the impurity (relative to the particle-hole symmetric point), 
$\hnd{n}_\sigma = \hnd{c}^\dagger_{0,\sigma} \hnd{c}_{0,
   \sigma}$ is the impurity density of spin $\sigma$ electrons.
$\hnd{c}^\dagger_{i,\sigma}$ and $\hnd{c}_{i,\sigma}$ are the creation and
annihilation operators for electrons on site $i$ with spin $\sigma$.
We use $\hbar = e = 1$.
$\theta(t)$ is the Heaviside function: We switch the interaction on at time $t=0$.  
Typical calculations will be performed for large
times so that the system has relaxed to its stationary regime.  The hopping
parameters are given by $\gamma_i = 1$ except for $\gamma_0=\gamma_{-1}=
\gamma$ which connect the impurity to the electrodes.  The calculations can be
performed for arbitrary values of $\gamma$. However, since we are not
interested in the large energy physics of the electrodes, we suppose that
$\gamma \ll 1$, \textit{i.e.} that the tunneling rate from the impurity to the
electrodes is energy independent $\Gamma = 2 \pi \gamma^2 \rho_F$ where
$\rho_F$ is the density of states of the electron reservoirs at the Fermi level.
The non-interacting retarded Green's function of the free impurity
is given by
\begin{equation}
    g^R(\omega) = \frac{1}{\omega - \epsilon_d + i\Gamma}.
\end{equation}
 The two electrodes have a chemical potential symmetric with respect to zero $\pm V_{\rm b}/2$ which 
 corresponds to a bias voltage $V_{\rm b}$. 
 They share the same temperature that we take very low $T=10^{-4}\Gamma$. Within the standard non-equilibrium Keldysh formalism \cite{Stefanucci2013}, the non-interacting
lesser and upper Green's functions are given by:
\begin{eqnarray}
g^<(\omega) &=& \frac{i\Gamma
\left[ n_F \left( \omega+\frac{V_{\rm b}}{2}\right ) + n_F\left (\omega-\frac{V_{\rm b}}{2}\right) \right]
}{(\omega - \epsilon_d)^2 + \Gamma^2},
\\
g^>(\omega) &=& \frac{i\Gamma
\left[  n_F \left(\omega+\frac{V_{\rm b}}{2}\right ) + n_F\left (\omega-\frac{V_{\rm b}}{2}\right) - 2\right]
}{(\omega - \epsilon_d)^2 + \Gamma^2},
\end{eqnarray}
where $n_F(\omega)=1/(e^{\omega/T} + 1)$ is the Fermi function.
$g^>(\omega)$ and $g^<(\omega)$ are the starting point for the expansion in power of $U$ that will be performed with real-time diagrammatic quantum Monte-Carlo.

The quantities of interest in this article are the interacting Green's functions (denoted with capital letters),
\begin{subequations}
\begin{eqnarray}
	G^R(t,t') &=& -i\theta(t-t') \left<\left\{\hnd{c}_{0\uparrow}(t), \hnd{c}_{0\uparrow}^\dagger(t')\right\}\right>, \\
    G^<(t,t') &=& i \left<\hnd{c}^\dagger_{0\uparrow}(t') \hnd{c}_{0\uparrow}(t)\right>, \\
    G^>(t,t') &=& -i \left<\hnd{c}_{0\uparrow}(t) \hnd{c}_{0\uparrow}^\dagger(t')\right>,
\end{eqnarray}
\end{subequations}
where the operators have been
written in Heisenberg representation. Since we will restrict ourselves to the
stationary limit, these functions are a function of $t-t'$ only and can be
studied in the frequency domain. Of particular interest is the spectral
function (or interacting local density of state) given by
\be
A(\omega) = -\frac{1}{\pi}{\rm Im}[G^R(\omega)].
\ee
The equilibrium spectral function displays the important features of Kondo physics: 
a sharp Kondo resonance at the Fermi level, and satellite peaks around
$\omega=\pm U/2$ in the case of particle-hole symmetry.

\begin{figure*}[ht]
    \centering
    \includegraphics[width=\textwidth]{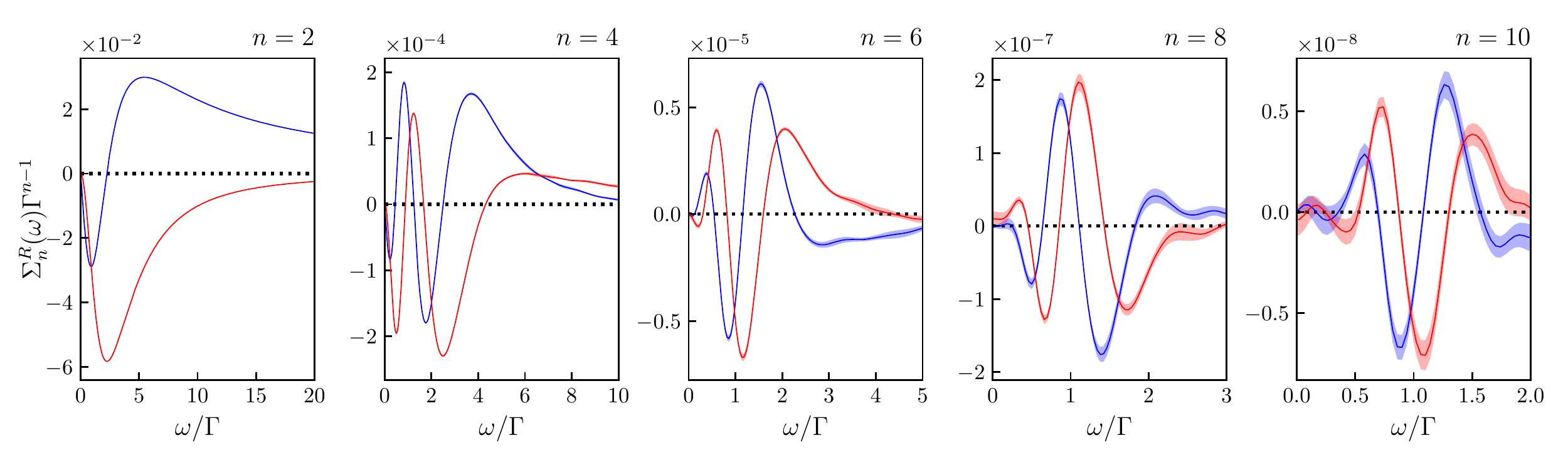}
    \caption{
        \label{fig:SE_order_by_order}
        First non zero orders of the self-energy series $\Sigma^R(U, \omega)$
        in powers of $U$ for the equilibrium particle-hole symmetric Anderson model (real
        part in blue, imaginary part in red). This series has been computed
        with a real-time diagrammatic quantum Monte-Carlo method detailed in a
        companion article\cite{Bertrand2019a}. The statistical error is shown as
        shaded areas. Due to particle-hole symmetry, odd orders are zeros.
    }
\end{figure*}

The out of equilibrium spectral function can be used for the computation of the current-voltage characteristic 
using the Wingreen-Meir formula\cite{Meir1992},
\be
\label{eq:landauer}
I = \frac{\Gamma}{2} \int  A(\omega) \left[ n_F \left(\omega+\frac{V_{\rm
b}}{2}\right) - n_F \left(\omega-\frac{V_{\rm b}}{2}\right) \right] d\omega.
\ee
The retarded self energy $\Sigma^R(\omega)$ is defined from the interacting
Green's function by:
\begin{equation}
\label{eq:self}
    G^R(\omega) = \frac{1}{\omega - \epsilon_d + i\Gamma - \Sigma^R(\omega)}.
\end{equation}
Physical quantities have systematic expansion in power of $U$
\be
G^R(t - t') = \sum_{n=0}^{+\infty} G^R_n(t-t') U^n,
\ee
from which we obtain the corresponding quantity in the frequency domain by Fourier transform,
\be
G^R(\omega) = \sum_{n=0}^{+\infty} G^R_n(\omega) U^n.
\ee
We obtain the functions $G^R_n(\omega)$ (typically up to $n=10$) using the QMC algorithm
of Ref.~\onlinecite{Bertrand2019a, Profumo2015}. The expansion of the self-energy 
\be
\Sigma^R(\omega) = \sum_{n=0}^{+\infty} \Sigma^R_n(\omega) U^n
\ee
is obtained from the $G^R_n(\omega)$
using a formal series expansion order by order of the Dyson equation (\ref{eq:self}). 
As an illustration, Fig.~\ref{fig:SE_order_by_order} shows the self-energy series, up to order $10$, for the equilibrium particle-hole symmetric model as obtained from diagrammatic QMC\cite{Bertrand2019a}. 
These series are the starting point of this paper, 
which is devoted to the resummation of the perturbative expansion for the Green's function and the self-energy beyond weak coupling.

\section{The perturbative series beyond the weak coupling regime}
\label{sec:resummation}

Diagrammatic Quantum Monte Carlo yields the first orders of the perturbation
expansion of physical quantities, with some error bars.
In weak coupling, we can directly sum this series and obtain the physical quantities
with a few orders. Beyond weak coupling however, we face a more complex problem.
For a given physical quantity $F$, we want to evaluate $F(U)$ from 
the first $N$ (typically $N\sim 10$) coefficients $F_0$, $F_1$,
$F_2$\dots $F_N$ of a series $F(U) \equiv \sum_{n=0}^\infty F_n U^n$.
In the following, $F$ will stand for the width of the Kondo peak, the Green's function $G$ or the self-energy $\Sigma$ of the impurity.
In the latter cases, the coefficients are functions of the frequencies, $G_n(\omega)$ and $\Sigma_n(\omega)$.
We also want to know, for a given physical quantity $F$ and interaction $U$, how many orders $N_0$
are needed to obtain $F(U)$ at a given precision.
Since the cost of the diagrammatic QMC approach is exponential in $N_0$, 
the answer to this question gives the ultimate limit of the method.

The mathematical problem of series resummation is a quite old topic, \textit{e.g.} 
Ref.~\onlinecite{Hardy1949}. Various techniques have been used in physics problems including Pad\'e approximants
\cite{Baker1996}, Lindel\"of extrapolation\cite{Lindeloef1905, VanHoucke2012} or Ces\`aro-Riesz technique\cite{Prokofev2008}.
In diagrammatic QMC, this is typically a post-processing step: the Monte-Carlo
produces the values of the various orders of the expansion, and 
one then attempts to sum the series to obtain the final result.
However, the situation is quite different if we want to use
such technique to solve quantum impurity models in the context of 
the quantum embedding methods like DMFT\cite{Georges1996}, 
or \textit{e.g.} Trilex\cite{Ayral2015, Ayral2017}.
Indeed, in such cases, the method require multiple solutions of impurity model
to solve their self-consistency loop.
Therefore, it is necessary to develop more robust methods to sum the perturbative 
series for impurity systems, which could be automatized.

In the cases considered in this paper (quantum
impurity models), and in general for lattice models at finite temperature (such
as the Hubbard model), the series for $F$ is expected to have a non-zero radius of
convergence $R_F$. Note that $R_F$ not only depends on the chosen physical
quantity $F$, but may also depend on frequency, voltage, temperature, \textit{etc}.
$R_F$ separates the {\it weak coupling regime} ($|U|<R_F$) from the {\it strong
coupling regime} ($|U|>R_F$).  At weak coupling, the truncated
series $\sum_{n=0}^N F_n U^n$ provides an accurate estimate of $F(U)$ and is
controlled exponentially with the number of coefficients $N$ (like a geometric
series since $F_n\sim (1/R_F)^n$).  At strong coupling however, this truncated
series diverges.
Note that in some problems like \textit{e.g.} the unitary fermionic gas, the series has a 
zero radius of convergence at zero temperature,
see \textit{e.g.} Ref.~\onlinecite{Rossi2018} for a recent example with diagrammatic QMC. 
We will not consider these cases in this paper, as they require other techniques as the ones presented here, 
\textit{e.g.} Borel summation techniques. 

In this paper, we consider the series summation problem with the angle
of reconstructing the function $F(U)$ in the complex $U$ plane.
The divergence of the series is due to the presence of singularities
in the complex $U$ plane, starting on the circle $|U|=R_F$.
The question is to reconstruct $F$ {\it beyond the radius of convergence}.

\subsection{General theory}
\label{sec:resummation_theory}

\subsubsection{Conformal transformation}

Conformal transformations can be used to deform
the complex plane and bring the point to be computed back into the convergence disk of a transformed series.
This technique was used a long time ago \textit{e.g.} in statistical physics \cite{Guttmann1989}. 
In a previous work\cite{Profumo2015}, some of us have shown that a simple conformal Euler transform allows 
to compute the density on the impurity up to $U= \infty$, at very low temperature, from the first 12 coefficients of the series.
However, this Euler transform is not always successful in resumming other quantities like the Green's function 
and the self-energy, and needs to be generalized.

Suppose that we aim at evaluating $F(U)$ at $U= U_0$ with $U_0$ real, positive
and $U_0>R_F$.
First, we assume a {\it separation property}, {\it i.e.} that we can find a simply connected domain delimited by a curve ${\cal C}$ 
containing 0 and $U_0$ but no singularities of the function $F$, as
illustrated in the lower left panel of Fig.~\ref{fig:intro}. The singularities of the function $F(U)$ will be located outside the domain ${\cal C}$.
We then proceed as follows:
\begin{itemize}

\item First, according to the Riemann mapping theorem, we can construct a biholomorphic change of variable $W(U)$ such that {\it i) } $W(0) =
0$, {\it ii) } it maps the interior of ${\cal C}$ into a disk $D_{\cal C}$ centered at $0$ in the $W$ plane
(see the lower right panel in Fig.~\ref{fig:intro}).
In practice, we seek $\cal C$ to separate the singularities from the half straight line of real positive $U$.
In the following, we will use two simple transformations, but in general we could use a Schwarz–Christoffel map if ${\cal C}$ is a polygon \cite{Driscoll2002}, 
composed with a M\"obius transformation of the disk to enforce {\it i) }.

\item Second, we form the series for the reciprocal function $U(W)$ of $W(U)$ which is defined term by term by the equation $U(W(U))=U$.
We then construct the series $\bar F(W) \equiv \sum_p \bar F_p W^p$  defined by the composition
$\bar F(W) = F(U(W))$.
Since $W(0)= 0$,
the first $N$ terms of $F(W)$ can be computed from the first $N$ terms of $F(U)$.

\item We evaluate the series $\bar F(W_0)$ at the point of interest $W_0= W(U_0)$. 
Indeed, by construction $W_0\in D_{\cal C}$ and, since $\bar F(W)$ is holomorphic in  $D_{\cal C}$, $D_{\cal C}$ is included in the convergence disk 
of the series $\bar F$. Hence the series $\bar F$ converges at $W_0$.
\end{itemize}

The result is independent of the choice of the domain ${\cal C}$ but 
the speed of convergence of the series for $\bar F(W_0)$ versus $N$ is not, since it is determined by the relative position
of $W_0$ compared to the radius of convergence $R_{\bar F}$ of $\bar F$, \textit{i.e.} 
$ \eta_{{\cal C}} \equiv |W_0/R_{\bar F}|$.
Therefore, there are ways to optimize the domain ${\cal C}$.
For example, we can not simply take a narrow domain close to the real axis, for the convergence in $W$ would
be really slow: we need to have $U_0$ and $0$ as ``far" as possible from the curve ${\cal C}$
(the precise meaning of ``far" being given by $\eta_{{\cal C}}$).
For each domain ${\cal C}$ satisfying the separation property, there is a minimum number of orders 
$N_{{\cal C}}$ needed to obtain the result at a given precision $\epsilon$.
There is therefore an optimal domain, which minimize $N_{{\cal C}}$ to $N_{opt} = \min_{{\cal C}} N_{{\cal C}}$.
This is the absolute minimum of orders needed to sum the series, and therefore 
determine {\it in fine} the complexity of the diagrammatic QMC algorithm.
Our next goal will be to approach such optimum.

Note that a failure of the separation assumption, \textit{i.e.} the choice of a domain
containing singularities, may result simply in the divergence of the series $\bar F$ at $W_0$,
hence a clear failure of the method rather than a wrong result.
Conversely the study of the convergence radius of the $\bar F(W)$
series provides direct information on the singularity free regions of the $U$ plane. Indeed, the region of the $U$ plane that maps towards the inside of the convergence radius of $\bar F(W)$ are singularity/branch cut free. 
Hence, using several conformal transforms, one may perform a step by step construction of the domain
${\cal C}$. 
Another note is that, as a consistency check, one can also check the stability of the final result upon small deformations of the domain (or the $W(U)$ function), 
as was discussed in details in Ref.~\onlinecite{Profumo2015} for the Euler transform.

The existence of the domain ${\cal C}$ and the transformation $W(U)$ has a
direct consequence on the algorithmic complexity of diagrammatic Quantum Monte-Carlo.
It was shown in Ref.~\onlinecite{Rossi2017a} that, 
for values of $U$ {\it inside} the convergence radius, connected diagrammatic quantum Monte-Carlo techniques
provide a systematic route for calculating
the many-body quantum problem in a computational time that only increases {\it
polynomially } with the requested precision.
The result also applies to the Keldysh diagrammatic QMC.
For completeness, the core of the argument is as follows: 
inside the radius of convergence $R$, the precision of a calculation
$\epsilon$ increases exponentially with the number of orders $N$ used $\epsilon
\sim (U/R)^N$. Hence, although the computational time $C$ increases
exponentially with $N$,  $ C\sim a^N$, 
the overall computational time scales as $C \sim (1/\epsilon)^{\log a/\log (R/U)}$, \textit{i.e.}
polynomially, see Ref.~\onlinecite{Rossi2017a} for a detailed analysis.
For a given $U_0$ and domain ${\cal C}$, we now have to sum the transformed
series $\bar F$ inside the radius of convergence. Hence the same argument also apply for this series, 
and therefore we conclude that, {\it even outside the disk of convergence}, 
we expect the algorithm to have a polynomial complexity as a function of the precision.
Let us emphasize however that this result is largely academic, since in practice the power law can be large.
Moreover, as we will discuss, for some physical quantities
the transformation to $W$ can lead to a dramatic increase of the noise
which induces a large computation time for a given precision.

\subsubsection{Location of singularities in the complex $U$ plane}

In order to choose ${\cal C}$ properly, we need to have some information on the location
of the singularities in the $U$ plane. In this paper, we use the following technique to approximately
locate the poles of $F(U)$ in the complex plane. 

\begin{itemize}
    \item We form an inverse of $F$ of the form  $K(U)=1/(F(U)+a)$ as a formal series (\textit{i.e.} order by order).
	 $a$ is a constant that we choose at our convenience. In order for the series $K(U)$ to exist, we must have $F_0 + a \neq 0$.
   \item We estimate the radii of convergence $R_F$ (resp. $R_{K}$) of $F$ (resp. $K$), 
         by plotting $|F_n|$ and $|{K}_n|$ versus $n$, and fitting the asymptote $|F_n|\sim (1/R_F)^n$
   \item In most of situations we found $R_F \neq R_{K}$. If not, we used a different $a$ so as to obtain $R_F \neq R_{K}$. Without loss of generality, let us assume that $R_{K}$ is the largest.
         We use the truncated polynomial of the series,  $\sum_{p=0}^N {K}_p U^p$
	 to compute $K(U)$ within its disk of convergence and therefore locate its zeros, which are the poles of $F$.
         They will appear as the accumulation of the zeros of the polynomials at large enough $N$.
	 If $R_F > R_{K}$, we simply reverse the roles of the series and reconstruct $K(U)$.
\end{itemize}

This technique has a quite large degree of generality, but also limitations.  It
assumes for example that the leading singularities in $F$ are poles and that the
radius of convergence of $F$ and $K$ are different. 
Also it does not give us indications of poles that would be far from the origin but close to the real axis.
However, in practice, we will see
below that for the quantities and the physical problem considered in this paper (Green's function and self-energy in
real frequency, and Kondo temperature), this technique is sufficient.
Finally, once $F(U)$ has been re-summed, it can be used to locate its zeros, hence for the resummation of $K(U)$
which provides another consistency check of the method.

\subsubsection{Controlling the noise amplification using non-perturbative information and Bayesian inference}
\label{sec:theory:Bayes}

The transformation from $F_n$ to $\bar F_p$ is a linear one (with a lower triangular matrix), for a given transformation $W(U)$.
Depending on the eigenvalues of the corresponding matrix, the Monte-Carlo error bar in $F_n$ may be strongly amplified by the transformation.
As a result, the method may become unusable at strong coupling, as will be
illustrated below on Fig.~\ref{fig:TK_bayesian_inference}.

However, if we add some {\it non-perturbative information}, such as the fact that the Kondo
temperature vanishes at infinite $U$, or a sum rule, we can
construct a Bayesian inference technique that may be used
to decrease the statistical uncertainty.
Bayesian inference provides a systematic and rigorous way to incorporate this information into
the results and improve their accuracy.
In the rest of this paragraph, we describe the general theory for this technique.
We will illustrate it in the following section.

Let us consider a series $F(U)=\sum_{n=0}^N F_n U^n$ where the $F_n$ are known with a finite precision.
We note $F = \{ F_0, F_1, \dots F_N\}$ the corresponding (vectorial) random
variable.
We calculate the mean values $\langle F_n\rangle$ and the corresponding errors $\delta_n$ within the
quantum Monte-Carlo technique. We assume that the coefficients $F_n$
are given by independent Gaussian variables. This forms the ``prior"
distribution $P_{\rm prior}(F=f)$ in the absence of additional information. 
\begin{equation}
P_{\rm prior}(F=f) = \prod_{i=0}^N \frac{1}{\sqrt{2\pi\delta_n}} e^{-\frac{(f_n-\langle F_n\rangle)^2}{2\delta_n^2}}
\end{equation}
Let us note the additional information $X$. $X$ is a random variable that can be
directly calculated from the series, $X=g(F)$ but whose actual value is also
known very precisely by other means. In the example below, $X$ will be the
value of $F(U)$ at large $U$. Bayesian inference amounts to replacing the prior
distribution with the posterior distribution $P(F=f| X=x_0)$ that incorporates
the knowledge of the actual value of $X$ (we note $P(A|B)$ the conditional
probability of event $A$ knowing event $B$). The value of $X$ is often known
exactly. However, due to the presence of truncation errors, its value cannot be enforced exactly,
and we suppose that it is known
with a small error $\varepsilon$. Eventually, we take the limit
$\varepsilon\rightarrow 0$. Hence, we assign to $X$ a Gaussian probability
distribution
$P_X(X=x) = 1/(\varepsilon\sqrt{2\pi}) \exp[-(x-x_0)^2/(2\varepsilon^2)]$ and define the posterior distribution as,
\begin{equation}
P_{\rm posterior}(F=f) \equiv \int dx P(F=f| X=x) P_X(X=x).
\end{equation}
Using Bayes formula $P(F=f| X=x) = P(X=x| F=f)P_{\rm prior}(F=f)/P_{\rm prior}(X=x)$ and the deterministic relation
$P(X=x| F=f)=\delta[x-g(f)]$, one arrives at,
\begin{equation}
\label{eq:bayes}
P_{\rm posterior}(F=f) = \frac{P_X(X = g(f))  P_{\rm prior}(F=f)}{P_{\rm
prior}(X=g(f))}.
\end{equation}
In practice, one proceeds as follows: (i) one generates many series according
to $P_{\rm prior}(F=f)$. We emphasize that these series result from a single
QMC run, hence are trivially generated (independent Gaussian numbers). Bayesian
inference implies no significant computational overhead (ii) One construct a
histogram of the values of $X$
to obtain $P_{\rm prior}(X=g(f))$. (iii) Each series is given a weight $P_X(X =
g(f))/P_{\rm prior}(X=g(f))$ which is used to calculate other observables such as
the value of $F(U)$ at different values of $U$. In
practice the results are insensitive to the choice of $\varepsilon$ as long as
it is chosen large enough so that a finite fraction of the sample contributes
to the final statistics.

\begin{figure}[ht]
    \centering
    \includegraphics[width=8cm]{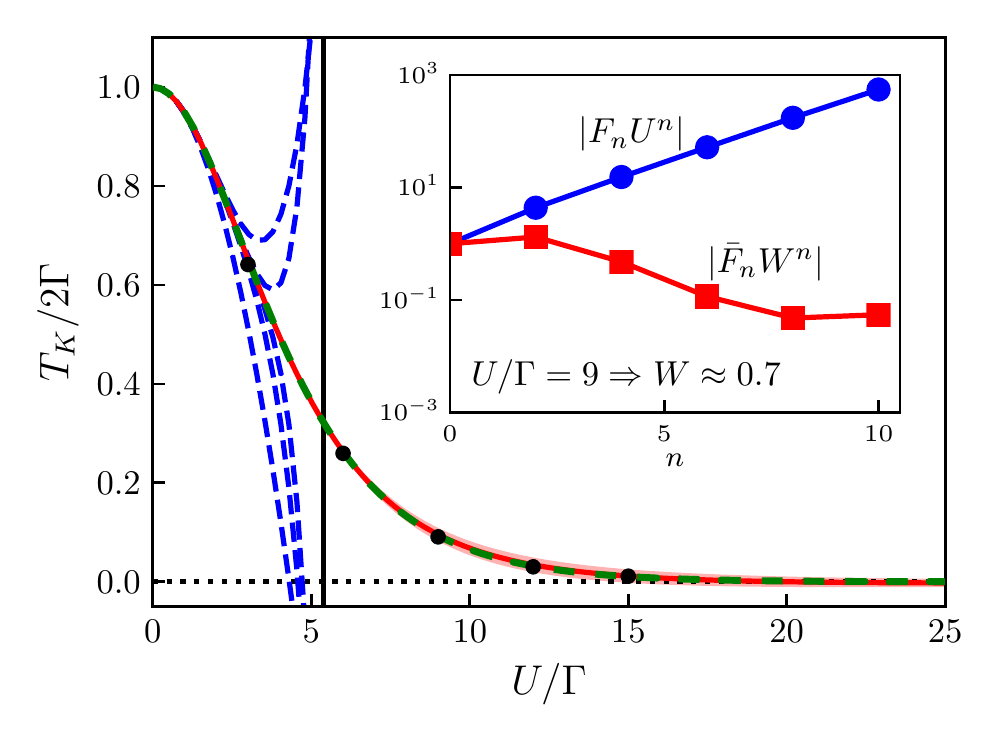}
    \caption{
    \label{fig:TK_resummed}
    Resummation of the Kondo temperature (as defined in Eq.~(\ref{eq:def_TK}))
    in the symmetric model ($\epsilon_d=0$).  Plain red line: resummation
    technique including Bayesian inference, using the Euler transform (error
    bar shown as red shaded area); dashed thick green line: exact result from Bethe ansatz\cite{Horvatic1985};
    black circles: reference NRG results; dashed blue lines: truncated series
    including up to $N=2,4,6,8$ and $10$ terms. The vertical line shows the
    estimated convergence radius of the series. Inset: evolution of $F_n U^n$
    with $n$ for $U=9\Gamma$ in log-linear scale (blue circles); evolution of
    the series $\bar F_n W^n$ obtained after conformal transformation (red
    squares). The value $W=0.7$ is obtained by applying the conformal
    transformation to $U=9\Gamma$. The $\bar F_n W^n$ decreases
    exponentially, indicating convergence of the transformed series while the
    original series (blue circles) diverges.
}
\end{figure}

\subsection{Illustration with the Kondo temperature}
\label{sec:resummation_kondo}

Let us first apply the method described above to the Kondo temperature $T_K$ (which will be $F$ in this section).
$T_K$ corresponds roughly to the width of the low energy Kondo peak,
and is defined more specifically in this paper as the dimensionful Fermi liquid quasi-particle weight extracted from the 
retarded self-energy at low energy:
\begin{equation}
    \label{eq:def_TK}
    T_K(U) \equiv \frac{2\Gamma}{1 - \left.\partial_\omega{\rm Re}\Sigma^R(U, \omega)\right|_{\omega=0}}.
\end{equation}
Our first goal is to illustrate how the method actually works, and benchmark it against
the calculation of the same quantity from the Numerical Renormalization Group (NRG) technique
and Bethe ansatz\cite{Horvatic1985}.

\begin{figure}[b]
    \centering
    \includegraphics[width=7cm]{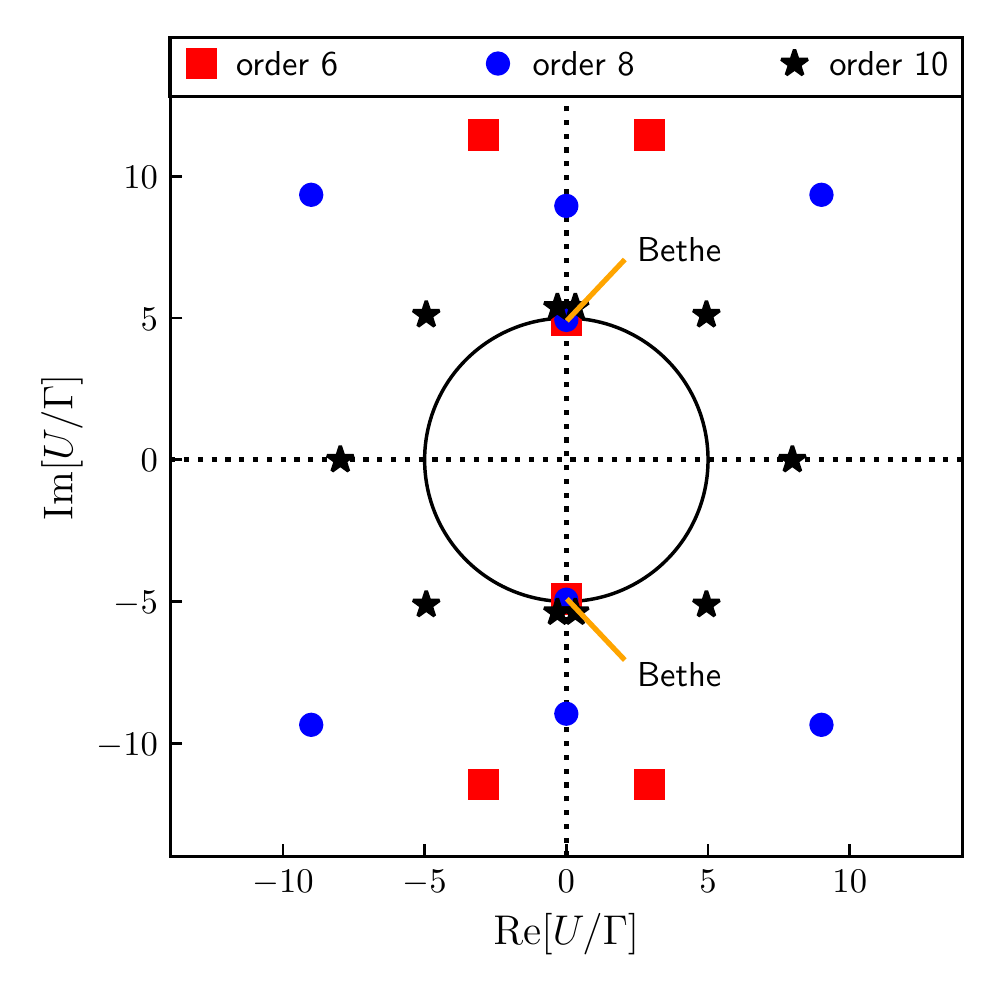}
    \caption{
        \label{fig:TK_poles}
    Poles of $T_K(U)$ identified from the zeros of the $1/T_K(U)$ function. 
    These are found by
    looking for the zeros of its truncated series.  Here they are shown in the
    $U/\Gamma$ complex plane with truncation at order 6 (red squares), 8 (blue
    points) and 10 (black stars). The black circle corresponds to $|U|=R_{T_K}$ where
    $R_{T_K}$ is the radius of convergence of the series of $T_K$. The stable 
    points close to $\pm i5\Gamma$ correspond to true non-perturbative poles of $T_K(U)$.   
    The exact zeros (small orange arrows) have been computed from the exact $1/T_K$
    series found with Bethe ansatz\cite{Horvatic1985}.
}
\end{figure}

\subsubsection{Singularities in the complex $U$ plane}

The dashed blue lines of Fig.~\ref{fig:TK_resummed} shows the truncated series of
$T_K$ = $\sum_{n=0}^N F_n U^n$ for various orders $N\le 10$.
These truncated series diverge around $R_{T_K}\approx 5\Gamma$ which is the convergence
radius of the series for these parameters. Increasing the value of $N$ helps to obtain a reliable
value of $T_K$ closer to $R_{T_K}$. However, as expected, even with a very large number of terms,
the bare series cannot be summed near or above $R_{T_K}$. Anticipating the final results, the plain red 
line corresponds to the results after resummation which matches very well what was obtained with our 
benchmark NRG calculation (see Sec.~\ref{sec:NRG} for details on the used NRG
implementation).

The inset of Fig.~\ref{fig:TK_resummed} shows the value of $|F_n U^n|$ (blue
circles) as a function of $n$ for $U/\Gamma=9$ which lies above the
convergence radius of the series. The log-linear plot shows an exponential
increase of $|F_n U^n| \sim (U/R_{T_K})^n$ with $n$ which we use to extract the
convergence radius of the series. Note that for other series, it can happen
that $|F_n|$ oscillates with $n$. Whenever $F_n$ changes sign, it becomes
close to zero which provides deviations from the clear exponential behaviour
shown in the inset of Fig.~\ref{fig:TK_resummed}. 
Hence, to obtain convergence radii which are robust to these outliers, we used a robust
regression method on the $\log|F_n|$ versus $n$ data (we compute the regression
slope as the median of all slopes between pairs of data points, this is known
in statistics as the Theil-Sen estimator\cite{Theil1992}).

We now compute the first $10$ terms of the series of $1/T_K(U)$. This series
has a radius of convergence of the order of $10\Gamma$. 
We look for the zeros, in the complex plane, of the series $1/T_K(U)$ truncated at order $N$.
Since the truncated series is a polynomial, it has
(generically) $N$ zeros, which are shown in Fig.~\ref{fig:TK_poles} for $N=6$
(red squares), $N=8$ (blue circles) and $N=10$ (stars). One pair of zeros $U
\approx \pm i 5\Gamma$ is converged for all the truncations, hence corresponds
to a true zero of $1/T_K(U)$, \textit{i.e.} to a pole of $T_K(U)$.
Fig.~\ref{fig:TK_poles} also shows the circle $|U|=R_{T_K}$ extracted from the
analysis of the $T_K(U)$ series done in the inset of
Fig.~\ref{fig:TK_resummed}. We find that the two poles $\pm i 5\Gamma$ do
indeed lie right on this circle. 

\begin{figure}[t]
    \centering
    \includegraphics[width=8cm]{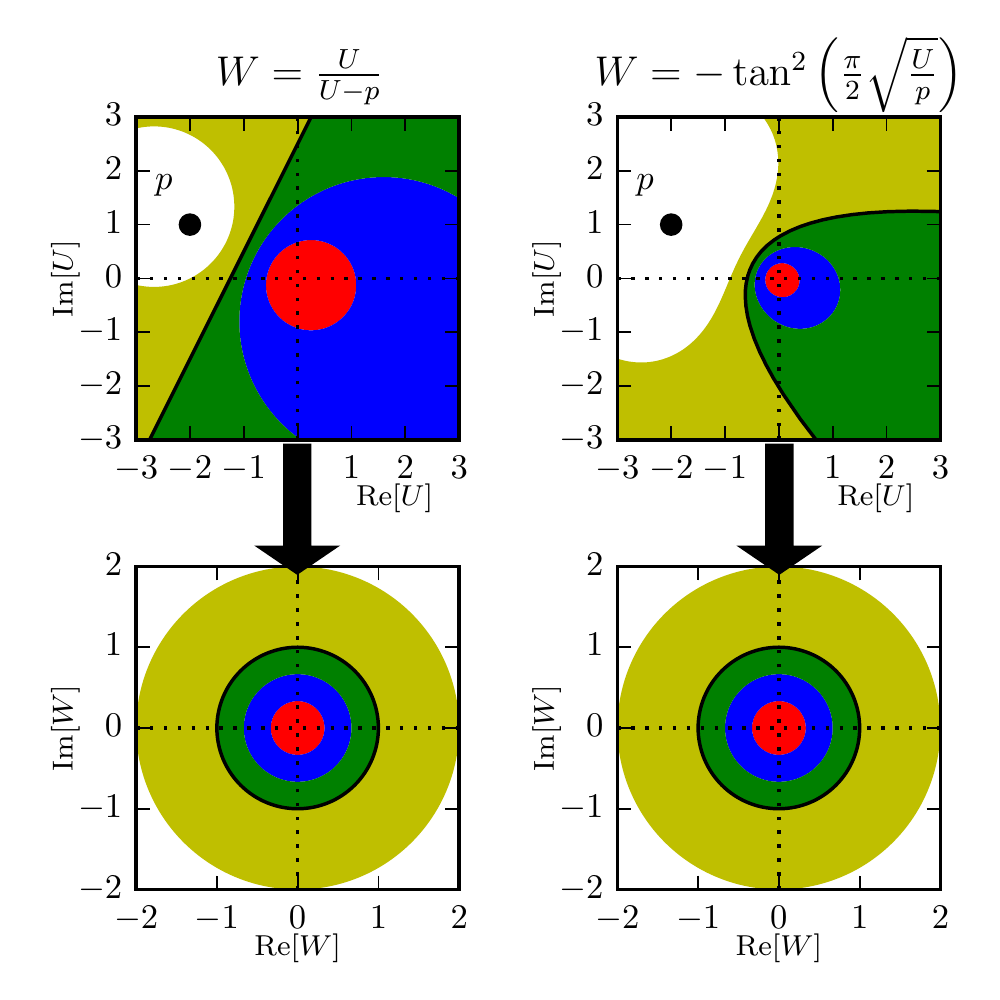}
    \caption{
        \label{fig:conformal_maps} Left panels: Euler map. Right panels: parabola map.
        Upper panels: complex $U$ plane. Lower panels: complex $W$ plane. The transformation
        maps the upper regions of various colors onto the lower regions of matching colors.
        In particular the thick straight line (upper left) and the parabola (upper right) are mapped onto the
        unit circles (lower left and right respectively).
    }
\end{figure}

\subsubsection{Conformal transformation}

Let us now turn to the conformal transformation $W(U)$, which maps the
two poles $\pm i5\Gamma$  away and brings the values of interest $U > 0$ (real)
closer to zero. 
We illustrate the technique with two maps:
the Euler map defined by 
\begin{equation}
    W = \frac{U}{U-p},
\end{equation}
and the ``parabola" map which is defined as
\begin{equation}
    W = - {\rm tan}^2 \left( \frac{\pi}{2} \sqrt{\frac{U}{p}} \right),
\end{equation}
where $p$ is an adjustable complex parameter.

Fig.~\ref{fig:conformal_maps} shows the various regions (different colors) in
the $U$ plane that are mapped onto concentric circles of the $W$ plane.  $0$ is
mapped onto $0$ and $p$ onto $\infty$ in both transforms. The Euler map (left
column) maps one half of the plane into the unit disk and the other half into
the outside of the unit disk (separated by a black line).  The parabola 
transform (right column) maps the inside of a parabola (black line) into the
unit disk and the outside of the parabola into the outside of the unit disk.
In the case where there are no singularities on the positive half plane ${\rm Re}[U]>0$,
the Euler transform should be preferred since real values of
$U>0$ are typically mapped closer to $U=0$ than with the parabola transform
(compare the size of the blue region of the parabola and Euler case for
instance).  However, the parabola map is more agnostic about the positions of
the singularities and will work even if there are singularities on the positive
half plane ${\rm Re}[U]>0$ as long as they lie outside the parabola.

We now perform the resummation of $T_K(U)$. The series contains only 
even power of $U$ due to particle-hole symmetry, so that it can be considered as
a function of $U^2$. The two poles $U=\pm i5\Gamma$ correspond to a
single one $U^2 = -25\Gamma^2$. In the $U^2$ plane, the pole being on the negative real axis,
the Euler maps works very effectively. The resummation can also be performed with the parabola transform.

Once the conformal map is selected, we form the series $\bar F_p$ in the $W$ variable, as explained above.
The inset of Fig.~\ref{fig:TK_resummed} shows $\bar F_n W_0^n$ (red squares) as a function of $n$ for
$W_0=0.7=W(U_0=9\Gamma)$, using the Euler map with $p=-35\Gamma^2$ (the parabola yields similar results with $p=-15\Gamma^2$).
As expected, $U_0$ is way beyond the radius of convergence in the original variable $U$, 
while $W_0$ lies within the disk of convergence of $\bar F(W)$ whose radius is found to be $R_{\bar F} \approx 2$.
The final result $T_K(U)$ using the Euler transforms is shown in
Fig.~\ref{fig:TK_resummed}. The parabola transform (not shown) is undistinguishable from
the Euler at this scale.

In this work, singularities were never found near the real positive axis, so
that all $U > 0$ can be reached using the conformal transforms of
Fig.~\ref{fig:conformal_maps}, given that enough orders of the series are
known. However, one may very well build a conformal transform to reach a regime
\textit{beyond} a singularity by considering a \textit{concave} contour $\cal
C$, as it is shown in Appendix~\ref{app:beyond_singularity}. This may become
interesting if a phase transition occurs when interaction is increased.

\begin{figure*}[t]
    \centering
    \includegraphics[width=\textwidth]{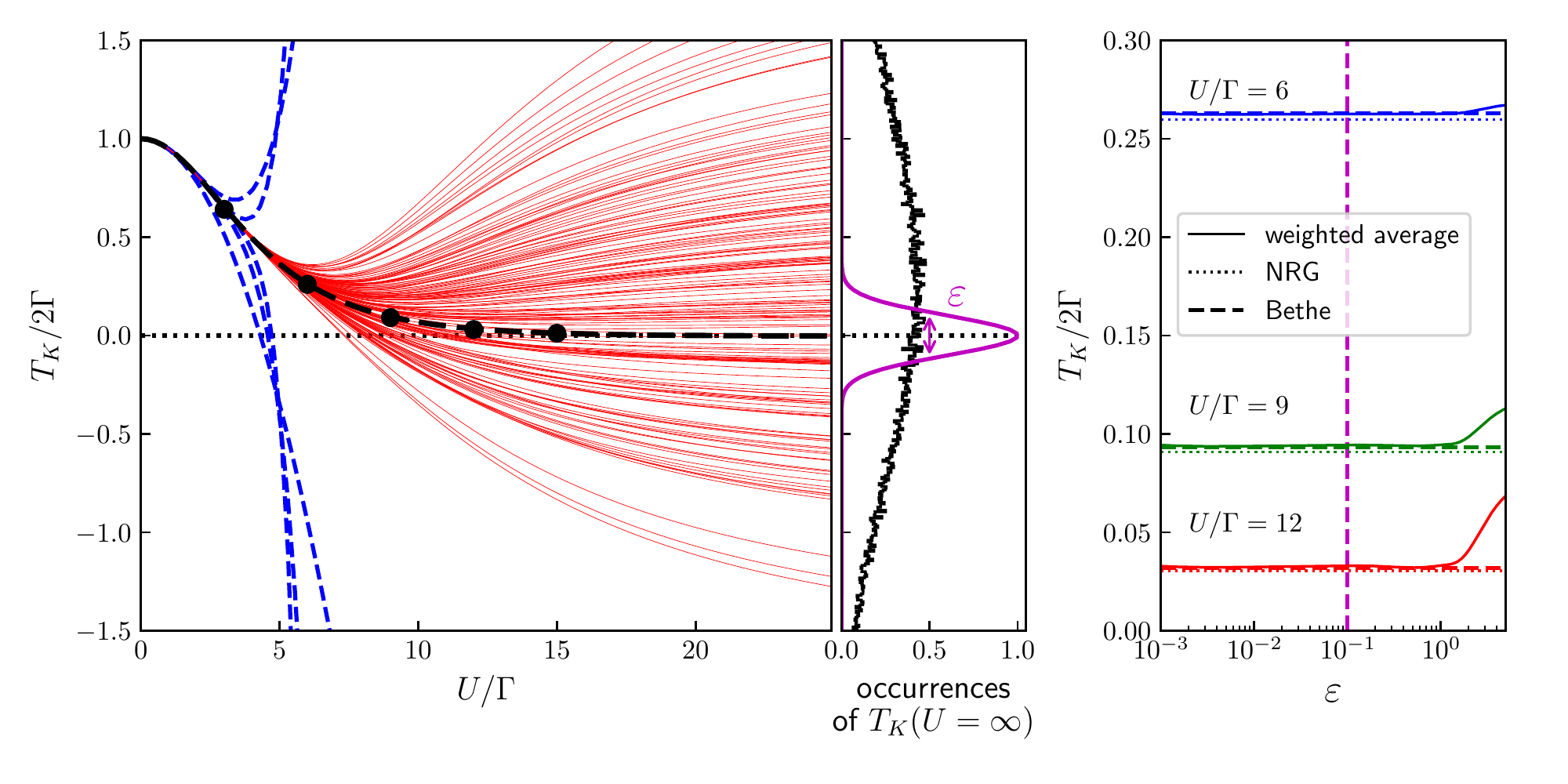}
    \caption{
        \label{fig:TK_bayesian_inference}
	Reduction of the statistical noise on the resummed $T_K(U)$ series by
	Bayesian inference.  Left panel: Kondo temperature as a function of
	$U$.  The bundle of red lines correspond to different samples of our
	series after resummation (see text).  The thick line shows the final
	result after Bayesian inference while the circles show our reference
	NRG calculations. The dashed blue lines show the bare results without
	resummation, which diverge for $U>5\Gamma$.  Middle panel: histogram of the
	values of $T_K(U=\infty)$ obtained from our samples (black line),
	histogram of its assumed distribution with tolerance $\varepsilon$
	(purple line). Right panel: final result after inference as a function
	of $\varepsilon$ for three values of $U/\Gamma=6,9$ and $12$ (thin lines),
	reference NRG result (dotted lines), Bethe ansatz result (dashed horizontal lines).
    }
\end{figure*}

\subsubsection{Noise reduction with Bayesian inference}

Let us now apply the Bayesian inference technique described above to the
computation of $T_K(U)$.
In the left panel of Fig.~\ref{fig:TK_bayesian_inference} we have
re-sampled the series for the Kondo temperature, \textit{i.e.} we have generated many
series (typically $10^3$ to $10^5$ samples).
For each sample we perform the
conformal transformation and plot the result for the Kondo temperature as a
function of $U$ (thin red lines). While we find that all results agree for $U
\le 6\Gamma$, the bundle of curves start to diverge for larger values of $U$. In the
middle panel, we plot (black thin line) the corresponding histogram of the
values obtained for $T_K(U=\infty)$, which is $P_{\rm prior}(T_K=g(f))$. 

We use the non-perturbative relation $\lim_{U\rightarrow\infty} T_K(U) = 0$.
Hence we want to ``post-select" the configuration of $F_n$ which 
give a vanishing Kondo temperature at large $U$, at precision $\epsilon$.
Following the  procedure described in Sec.~\ref{sec:theory:Bayes}, our final
result is obtained by averaging the different traces (thin red lines) with the
weight given by Eq.~(\ref{eq:bayes}). The right panel of
Fig.~\ref{fig:TK_bayesian_inference} shows the result for three different
values of $U$ as a function of $\varepsilon$ which confirms that the results
are insensitive to the actual value of $\varepsilon$. We find a very good
agreement with the results obtained from NRG even at large values of $U$, 
noting that NRG spectra have typical relative error bars of a few percents 
(see Sec.~\ref{sec:NRG} for details).

\subsubsection{Benchmark with the Bethe Ansatz exact solution}

The series expansion for $1/T_K(U)$ has been calculated explicitly and exactly
using the Bethe Ansatz technique by Horvatic and Zlatic\cite{Horvatic1985}.
Ref.\onlinecite{Horvatic1985} provides an iterative formula for calculating the
coefficients of the expansions and shows that the corresponding series has an
infinite radius of convergence.  This provide another independent benchmark of
the calculation of $T_K(U)$ as well as of the method itself. We checked that
the 10 first coefficients of this series agree with the one that we computed
with QMC.

Fig.~\ref{fig:TK_resummed} shows our final result together with the NRG result
(black circles) and the Bethe ansatz results. At this scale, the agreement is
perfect. Using the exact series for $1/T_K(U)$ (truncated to around 50
coefficients), we studied its zeros which are the poles of $T_K(U)$. We find
that they are situated on the imaginary axis. The poles closest to the origin
are $U/\Gamma \approx \pm 4.89059579i$ in agreement with our findings, see Fig.
\ref{fig:TK_poles}. The next poles are $U/\Gamma \approx \pm 13.79i$, $21.77i$,
$29.89i$, $37.87i$ and $45.9i$ but are too far to be accessible with only the
first ten coefficients. 
The right panel of Fig.\ref{fig:TK_bayesian_inference} provides a detailed benchmark of our
results versus both NRG and the exact Bethe Ansatz solution.

We find that the QMC results for $T_K$ are slightly more accurate than NRG, because
the extraction of $T_K$ from the NRG self-energy (see Eq.~(\ref{eq:def_TK})) contains
inherent broadening errors. The agreement between all three methods is nevertheless
excellent. In addition, we can extract from the Bethe Ansatz the exact QMC
error, and this error matches the measured 1 sigma statistical error bars.

\begin{figure}[t]
    \centering
    \includegraphics[width=8cm]{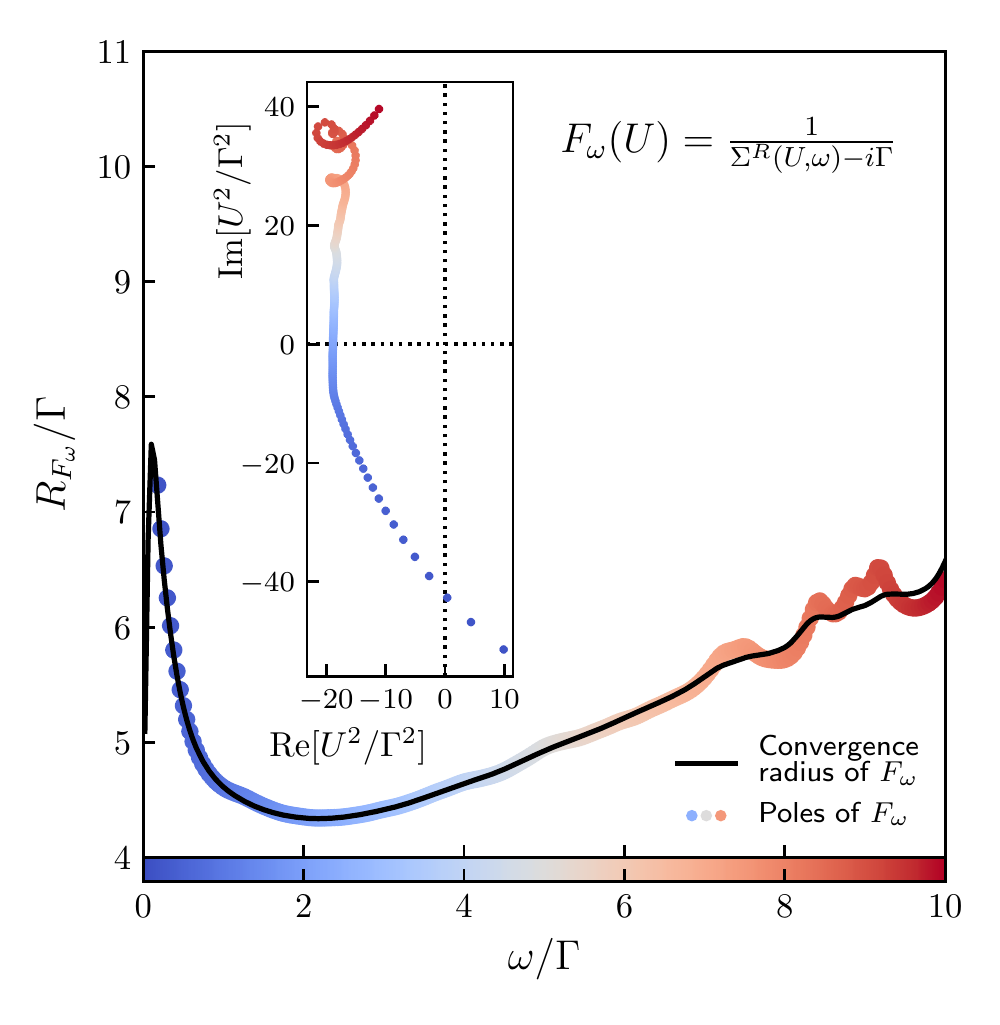}
    \caption{
    \label{fig:phsym_SE_zeros}
    Main frame: convergence radius $R_{F_{\omega}}$ of $F_\omega(U)=1/(\Sigma^R(U, \omega)-i\Gamma)$ (thin line) in the equilibrium symmetric Anderson
    impurity model. The color circles show the absolute value of the pole of $F_\omega(U)$.
    Inset: position of the pole of $F_\omega(U)$ in the $U^2$ complex plane for different frequencies.
    The color blue to red corresponds to increasing frequency, as in the main frame.
    At high frequency, the statistical uncertainty prevents an accurate localization of the poles.	
    }
\end{figure}

\subsection{Equilibrium dynamical correlation functions}
\label{sec:resummation_gf}

Let us now apply our method to the 
Green's function and self-energy as a function of the real frequency $\omega$.

\subsubsection{Singularities in the long time (stationary) limit}

Let us now turn to the full Green's function $G^R(\omega,U)$ and self-energy
$\Sigma^R(\omega,U)$. 
An example of our bare data is shown in Fig.~\ref{fig:SE_order_by_order} where we plot the coefficients
$\Sigma_n^R(\omega)$ obtained from real time diagrammatic quantum Monte-Carlo for $n=2,4,6,8$ and $10$.
The description of the method used to calculate these coefficients $\Sigma_n^R(\omega)$ is explained in the companion paper to this
article\cite{Bertrand2019a}.

We focus on the quantity
$\Sigma^R(\omega)-i\Gamma$ and denote its inverse
$F_\omega(U)=1/(\Sigma^R(\omega)-i\Gamma)$. The retarded Green's function can
be recovered from $F_\omega(U)$ using $G^R(\omega) = 1/(\omega
-F_\omega(U)^{-1})$ (using $\omega -\Sigma^R(\omega) +i\Gamma$ turns out to be less 
convenient especially at high frequency).

Fig.~\ref{fig:phsym_SE_zeros} shows the convergence radius of $F_\omega(U)$ as
a function of frequency, extracted from a study of the exponential decay of the
corresponding series with $n$. We have also performed a systematic study of the
zeros of $\Sigma^R(\omega)-i\Gamma$ in order to localize the poles of
$F_\omega(U)$. We find one pair of poles at each frequency. The results are
shown in the inset of Fig.~\ref{fig:phsym_SE_zeros} for a set of frequencies
from $\omega=0$ to $\omega=10\Gamma$ in the complex plane for $U^2$.  The
absolute value of the poles of $F_\omega(U)$ is also plotted in the main frame
of Fig.~\ref{fig:phsym_SE_zeros} as a function of frequency (circles of varying
colors from blue to red). We observe a perfect match with our estimation of the
convergence radius reflecting the fact that these poles are responsible for the
divergence of the series. It is important to note here that working in the
real frequency domain is very helpful: we found a single pole per
frequency (at least for the range of interactions that we could study). 
Hence, we expect that performing the resummation in real time or imaginary frequencies
could be more complex, since all these poles would be involved simultaneously.

\begin{figure}[t]
    \centering
    \includegraphics[width=8cm]{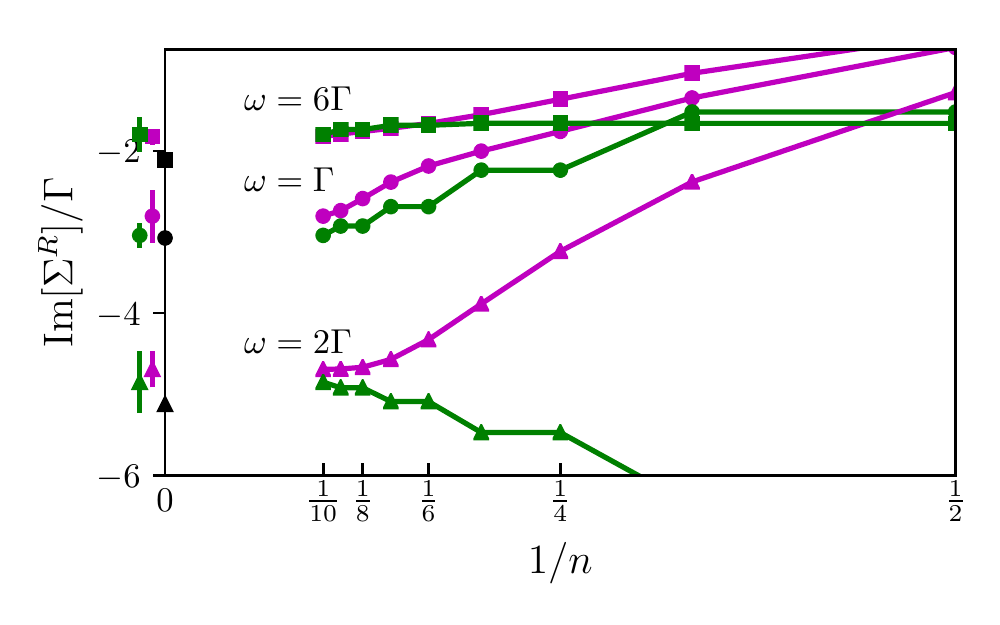}
    \caption{
        \label{fig:SE_resummed_details}
    Resummation of the self-energy in the equilibrium symmetric Anderson
    impurity model at $U=9\Gamma$.
    The imaginary part of $\Sigma^R(\omega)$ is shown as a function of the number $n$ of terms kept in the resummation, for three frequencies $\omega = \Gamma$ (circles), $2\Gamma$ (triangles) and $6\Gamma$ (squares).
    The independent resummation of $F_\omega(U)$ (green line) and of $\Sigma^R(\omega)-i\Gamma$ (purple line) converge with one another.
    The results with truncation and statistical errors are shown on the left of the y-axis, along with NRG results (black symbols).
    }
\end{figure}

The results for three frequencies ($\omega/\Gamma = 1, 2$ and $6$) are given in 
Fig.~\ref{fig:SE_resummed_details}. We show the convergence of the imaginary
part of the self-energy using two different resummed series: $F_\omega(U)$
(green symbols) and $1/F_\omega(U)$ (purple symbols).
The former has been resummed with an Euler transform with a frequency
dependant $p$ set close to the poles shown in Fig.~\ref{fig:phsym_SE_zeros}.
The latter, for which our method did not detect poles, has been resummed with
the parabola transform (in the $U$ plane) with $p=-4.5\Gamma$. Again, Bayesian
inference has been used to enforce $\lim_{U\rightarrow \infty} G(U, \omega) = 0$ for all $\omega \neq 0$.
For comparison, we also include the NRG results (which are
very accurate at small frequency and possibly less accurate at large frequency). 
 The slight difference between the purple and
green curves is due to the truncation error.  We find
that the series which has (initially) the largest convergence radius is less
sensitive to truncation error or statistical noise than the other. 
We attribute the small discrepancy
between the QMC results and NRG at large frequency to a lack of convergence of
the latter. These results are obtained for a rather strong interaction
$U=9\Gamma$. At smaller interaction the QMC and NRG results become
undistinguishable. At larger interactions, the QMC results become increasingly
inaccurate due to truncation errors.   

\begin{figure}[t]
    \centering
    \includegraphics[width=8cm]{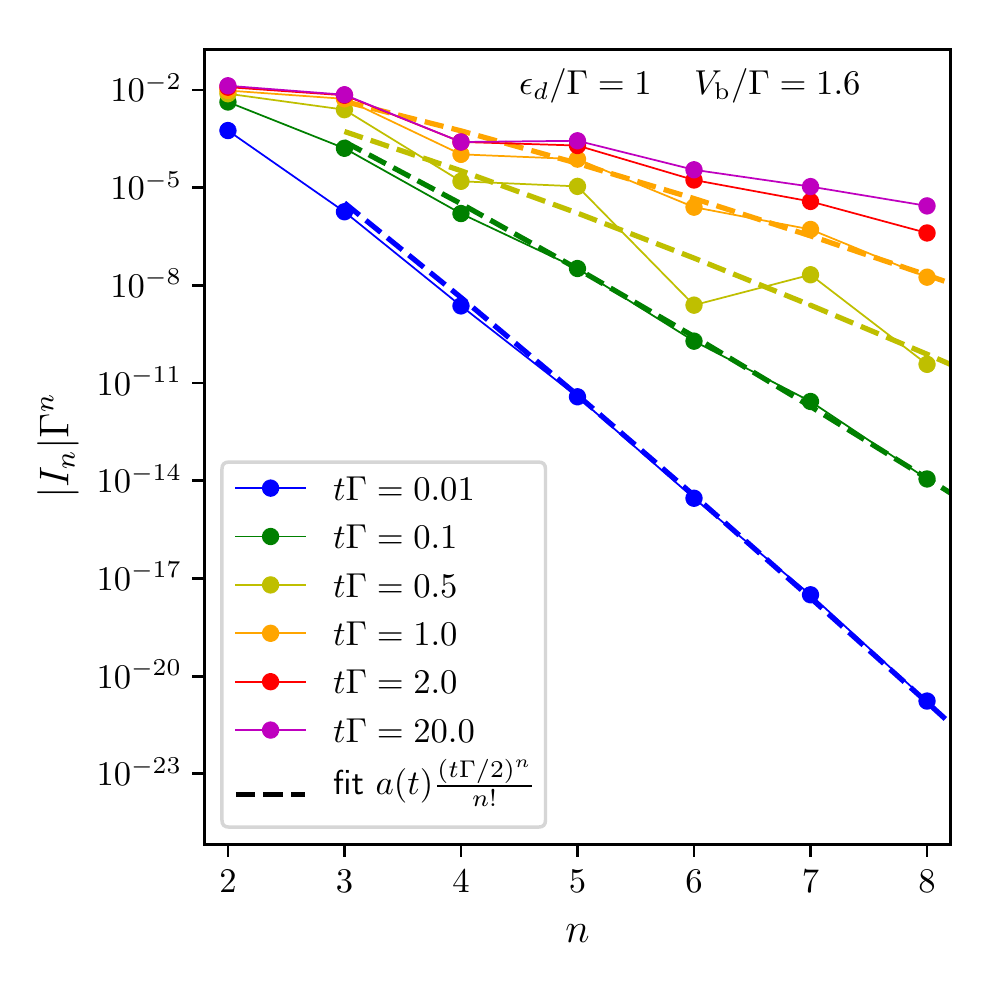}
    \caption{
    \label{fig:infinite_conv_rad}
    Coefficients (absolute value) of the series for the current (circles with thin lines) in the asymmetric model ($\varepsilon_d = \Gamma$, $V_{\rm b} = 1.6\Gamma$) computed at different times $t$ (different colors).
    The apparent convergence radius decreases with time.
    For small values of $t$, we can observe that the series coefficients decrease faster than exponentially, which indicates an infinite convergence radius.
    The thick dashed line shows the corresponding fit with $(t\Gamma/2)^n / n!$.
    For large enough $t$, the series converges toward the steady state limit.
    }
\end{figure}

\subsubsection{The long time limit}

In the Keldysh formalism, the interactions are switched on at an initial time (0),
and one follows the evolution of the system with time $t$.
We assume here that the system relaxes to a steady state at long time.
Let us consider the average of an operator ${\cal \hat O}$ as a function of time, and 
its expansion $\langle{\cal \hat O}(t) \rangle = \sum_n O_n(t)U^n$ (the extension of the following arguments 
to Green's function is straightforward).

At finite time $t$, the radius of convergence of this series is infinite, as shown in 
Appendix \ref{app:convergence_finite_time}.
Each order in the perturbation expansion $O_n(t)$ relaxes with $t$ to a long time limit, 
but the time $t_\text{relax}(n)$ it takes to reach this limit can increase with $n$.
The long time and large $n$ limit do not commute in general:
\begin{equation}
\lim_{n\rightarrow\infty} \lim_{t\rightarrow\infty}O_n(t) \neq
\lim_{t\rightarrow\infty} \lim_{n\rightarrow\infty} O_n(t).
\end{equation}
This behaviour was already noted in Fig.~14 of Ref.~\onlinecite{Profumo2015}.
It is also illustrated on Fig.~\ref{fig:infinite_conv_rad}, which shows 
various orders $n$ of the expansion of the current through the dot versus $n$, 
for different times.
We observe that at small times the orders $I_n$ decreases faster than exponentially with $n$, 
consistent with the bound mentioned above. The coefficients converge
to the steady state limit at long time.

At finite time $t$, since the series converges, it is sufficient to have enough orders.
In the steady state, as explained above, we have a minimal order $N_0$ needed
to compute the quantity at a given precision. One should then simply compute at a time
$t > t_\text{relax}(N_0)$.

In the Anderson model, some quantities like the spectral function are known to relax
on a long time scale $t_K \sim T_K^{-1}$, see \textit{e.g.} Ref.~\onlinecite{Nordlander1999}.
The previous remarks explain how the algorithm deals with this long time.
For a given $U$, we need $N_0(U)$ orders, hence to compute at a time larger than $t_\text{relax}(N_0(U))$.
The larger $U$ is, the longer this time becomes. However, it is still finite at fixed $U$, 
and since our calculation of the perturbative expansion is {\it uniform} in time, 
it is not an issue (the computation effort does not grow with time).
However, the existence of the Kondo time indicates that the number of orders necessary to compute \textit{e.g.} the low frequency spectral function
at a given $U$ increases with $U$ (otherwise the relaxation time of the physical quantity would be bounded at large $U$).

\section{Benchmark of the dynamics in equilibrium}
\label{sec:bench}

We now benchmark our results in the case of equilibrium, testing 
various regimes of the Anderson impurity model.
Let us first describe the high-precision NRG computations that were performed.

\subsection{NRG implementation}
\label{sec:NRG}

The Numerical Renormalization Group (NRG)~\cite{Bulla2008}
was used to benchmark our QMC calculations in equilibrium,
and to test the reliability of the series extrapolation method
for spectral functions at various values of $U$ and $\epsilon_d$.
In order to obtain precise NRG data for the spectral function 
of the Anderson impurity model, the computations were performed 
using several improvements over the simplest implementations of
the NRG. First, the full density matrix formulation of 
NRG~\cite{Hofstetter2000} was used to reduce finite size effects
due to the NRG truncation. Second, symmetries of the problem were heavily 
exploited~\cite{Toth2008}, allowing to reduce significantly the Hilbert 
space dimension of various multiplets. In the particle-hole symmetric case, 
the full SU(2)$_\mathrm{charge}\otimes$SU(2)$_\mathrm{spin}$
symmetry was used, while the charge sector was reduced to U(1)$_\mathrm{charge}$ 
away from particle-hole symmetry. Third, the impurity Green's function
was extracted from a direct computation of the $d$-level
self-energy $\Sigma(\omega)$~\cite{Bulla1998}, according to its exact 
representation as the ratio of two retarded correlation functions in the
frequency domain:
\begin{equation}
\Sigma(\omega) = U \frac{F^R(\omega)}{G^R(\omega)},
\end{equation}
where $G^R(t)=-i\theta(t)\langle
\{d^{\phantom{\dagger}}_\sigma(0),d_\sigma^\dagger(t)\}\rangle$ is the
usual single particle retarded Green's function in the time domain, and
$F^R(t)=-i\theta(t)\langle \{d^{\phantom{\dagger}}_\sigma(0) 
d_{-\sigma}^\dagger(0) d^{\phantom{\dagger}}_{-\sigma}(0),
d_{\sigma}^\dagger(t) \}\rangle$ is a composite fermionic correlation function.
In practice, $\mathrm{Im}[G^R(\omega)]$ and $\mathrm{Im}[F^R(\omega)]$ are
computed from the K\"all\'en-Lehmann representation using the broadened NRG 
spectra, and the real parts of both $G^R(\omega)$ and $F^R(\omega)$ are obtained 
via a Kramers-Kronig relation.
Finally, the truncation parameters of the NRG simulations were taken 
to model as closely as possible a continuous density of states for
the electronic bath. Although the use of the logarithmic Wilson
discretization grid, $\omega_n = D \Lambda^{-n}$, is inherent to 
the practical success of NRG, we found that values of $\Lambda$
as low $\Lambda=1.4$ could be managed in practice within the NRG, 
taking a very large number $N_\mathrm{kept}=3200$ of kept multiplets.
Up to $N_\mathrm{iter}=120$ NRG iterations were used, so that the
effective temperature can be considered to be practically zero.
With such small value of $\Lambda$, the broadening parameter $b$ 
of the discrete NRG spectra could be decreased down to $b=0.2$, without
$z$-averaging, which further enhanced the spectral resolution of 
the Hubbard satellites in the spectral function.

\begin{figure}[t]
    \centering
    \includegraphics[width=8cm]{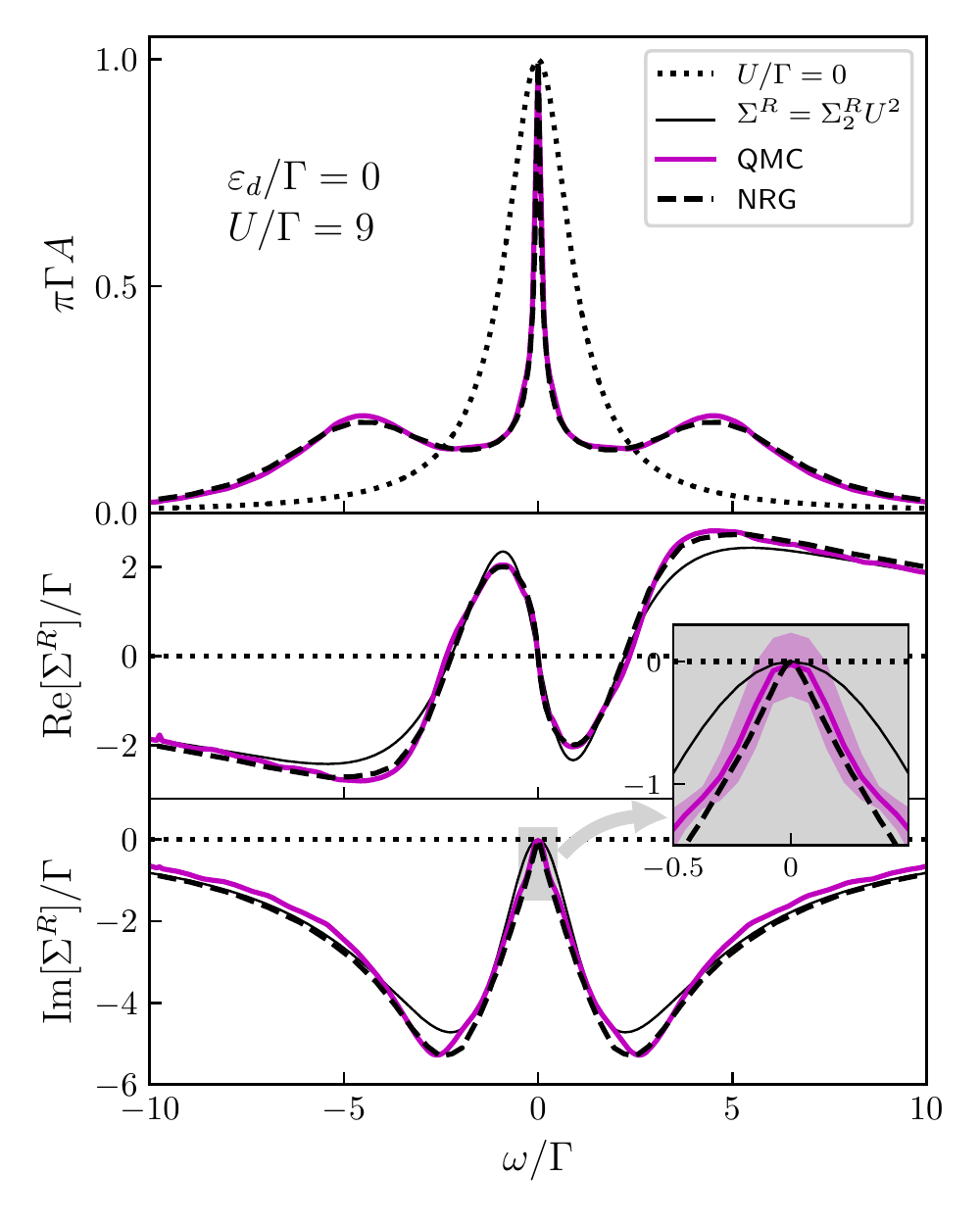}
    \caption{
    \label{fig:resummed_epsd_0}
    Resummed equilibrium spectral function (upper panel), real part (middle
    panel) and imaginary part (lower panel) of the retarded self energy
    $\Sigma^R(\omega)$ for the symmetric Anderson impurity $\epsilon_d/\Gamma =
    0$ at $U=9\Gamma$.
Purple line: resummed result from $10$ orders of perturbation theory; dashed
line: NRG; dotted line: non-interacting result; thin black line: second order
perturbation theory for the self-energy. Inset: zoom of the imaginary part at
small energy with error bars.
    }
\end{figure}

\begin{figure}[t]
    \centering
    \includegraphics[width=8cm]{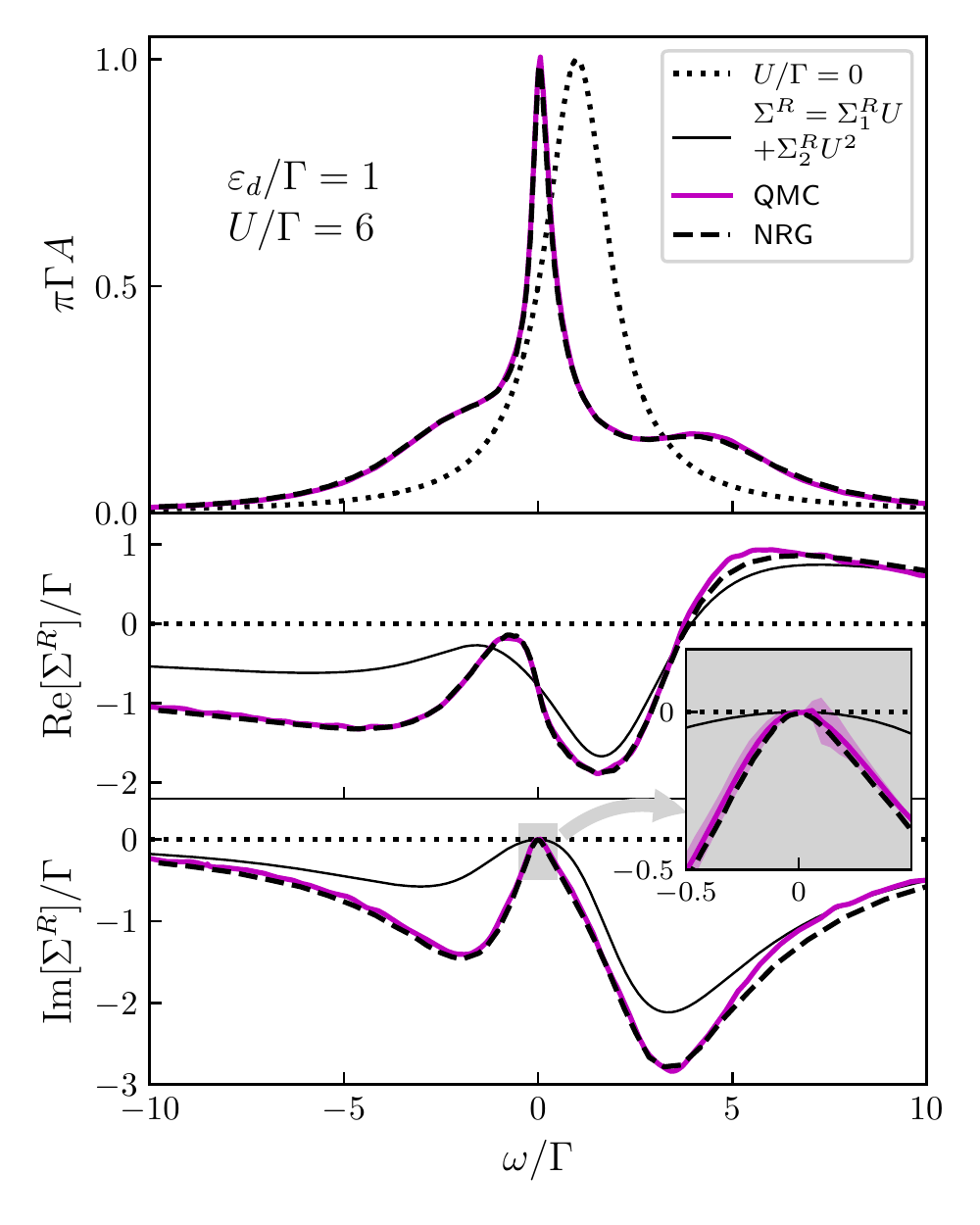}
    \caption{
    \label{fig:resummed_epsd_1}
    Resummed equilibrium spectral function (upper panel), real part (middle
    panel) and imaginary part (lower panel) of the retarded self energy
    $\Sigma^R(\omega)$ for the asymmetric Anderson impurity $\epsilon_d/\Gamma
    = 1$ at $U=6\Gamma$.  Purple line: resummed result from $10$ orders of
    perturbation theory; dashed line: NRG; dotted line: non-interacting result;
    thin black line: second order perturbation theory for the self-energy.
    Inset: zoom of the imaginary part at small energy with error bars.
    }
\end{figure}

%
\subsection{Comparison to NRG in equilibrium}
\label{sec:resultsEq}

Fig.~\ref{fig:resummed_epsd_0} shows the spectral function as well as the imaginary and real part of the
self energy for the symmetric Anderson impurity in the strong correlation
regime $U=9\Gamma$ (same data as the purple curve of Fig.~\ref{fig:SE_resummed_details}). 
The spectral function shows a clear Kondo peak and the two satellites at
$\omega \simeq \pm 4.5\Gamma=\pm U/2$ in good agreement with the NRG data. For this calculation, 
a simple second order calculation of the self-energy already provides a reasonably good result 
(thin black line), due to near cancellations in higher order diagrams in
the peculiar case of particle-hole symmetry. 

Fig.~\ref{fig:resummed_epsd_1} shows the same plot in the asymmetric case
$\epsilon_d = 1$. This case is more complex because the resonance at $U=0$ is
offset with respect to the Fermi level, hence to the position of the Kondo peak. We note
that previous real time QMC techniques suffered from a strong sign problem and
could not access the asymmetric regime \cite{Werner2010}.  We also stress that the
second order approximation is now very different from the correct result.
The comparison to the NRG data is still excellent.

\begin{figure}[t]
    \centering
    \includegraphics[width=8cm]{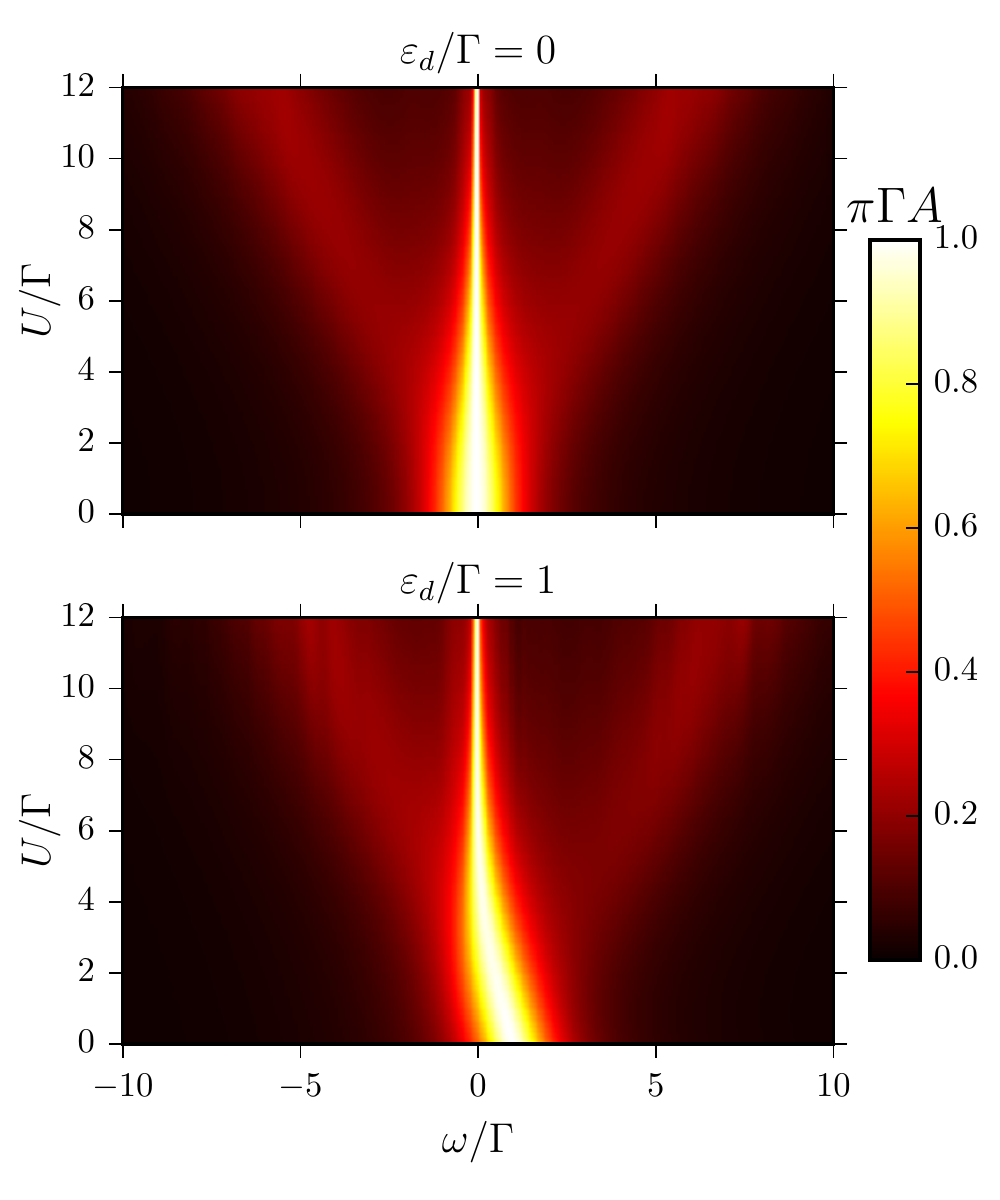}
    \caption{
    \label{fig:SF_colormaps}
    Color plot of the spectral density $A(\omega, U)$ in the symmetric case ($\epsilon_d/\Gamma=0$, upper
    panel) and asymmetric case ($\epsilon_d/\Gamma=1$, lower panel) as a function of $\omega$ and $U$.
    The data from each panel has been obtained in a single QMC run.
    }
\end{figure}

Another advantage of the techniques described in this article and its
companion article \cite{Bertrand2019a} is that a single QMC run provides the full
dependence in both $\omega$ and $U$, which is very time consuming in the NRG. 
This is illustrated in
Fig.~\ref{fig:SF_colormaps} where the color map shows the spectral function as a
function of $\omega$ and $U$. One can clearly observe the formation of the
Kondo peak (which gets thinner as one increases $U$ and shifts toward
$\omega=0$ in the asymmetric case) as well as the Hubbard bands at $\omega=\pm
U/2$. Note that the results are perfectly well behaved (qualitatively correct)
up to very large $U$ (even above $U=12\Gamma$ shown in
the plot) but  become quantitatively inaccurate at too large values of $U$.
Improving them would require the use of higher perturbation orders.

\begin{figure}[t]
    \centering
    \includegraphics[width=8cm]{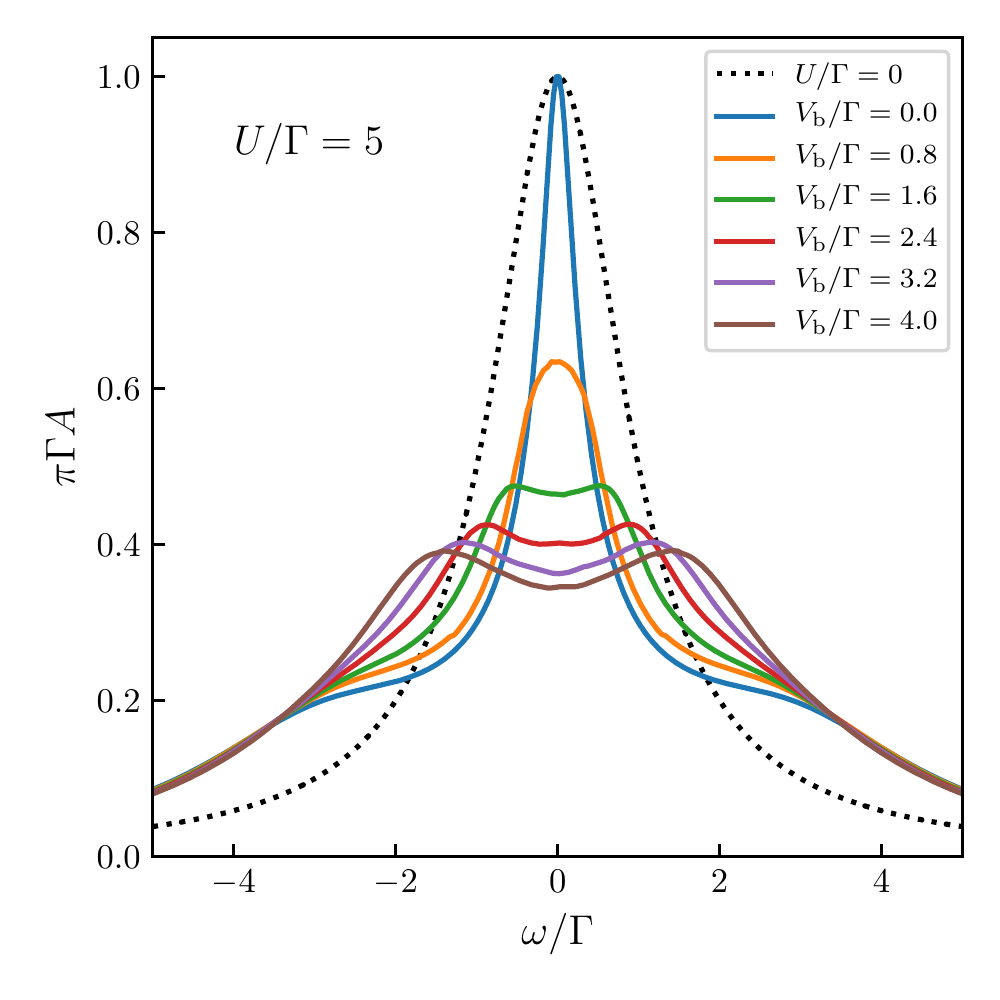}
    \caption{
    \label{fig:SF_sym_bias}
    Out-of-equilibrium spectral functions with interaction strength
    $U/\Gamma=5$, in the symmetric ($\epsilon_d/\Gamma=0$) model with a symmetric voltage bias
    $V_{\rm b}$. The resulting self-energy series has
    been resummed in a similar fashion as for the previous results. The
    non-interacting spectral function is shown as a dotted line.
    }
\end{figure}

\section{Out of equilibrium results}
\label{sec:results}

We finally turn to the out-of-equilibrium regime, and present some 
accurate computation of current-voltage characteristics, as well as 
novel predictions for dynamical observables in presence of a finite
bias voltage.

\subsection{Splitting of the spectral function}
\label{sec:resultsNonEq}

Fig.~\ref{fig:SF_sym_bias} shows
the spectral function of the symmetric impurity in the presence of various bias
voltages from $V_{\rm b}=0$ to $4\Gamma$. The results were obtained using
the parabolic map on the series of $\Sigma_\omega(U^2) - i\Gamma$ (with an
optimized frequency dependent parameter  $p/\Gamma^2\in[-25, -200]$). Upon
increasing the bias voltage, we find as expected from NCA\cite{Wingreen1994}
and perturbative\cite{Fujii2003} calculations that the Kondo resonance
simultaneously broadens and get split into two peaks.  Previous results on
the spectral function\cite{Cohen2014b} were based on the bold diagrammatic
approach and were calculated at relatively high temperature ($T = \Gamma/3$)
while using a third terminal for computing the spectral function.

\begin{figure}[t]
    \centering
    \includegraphics[width=8cm]{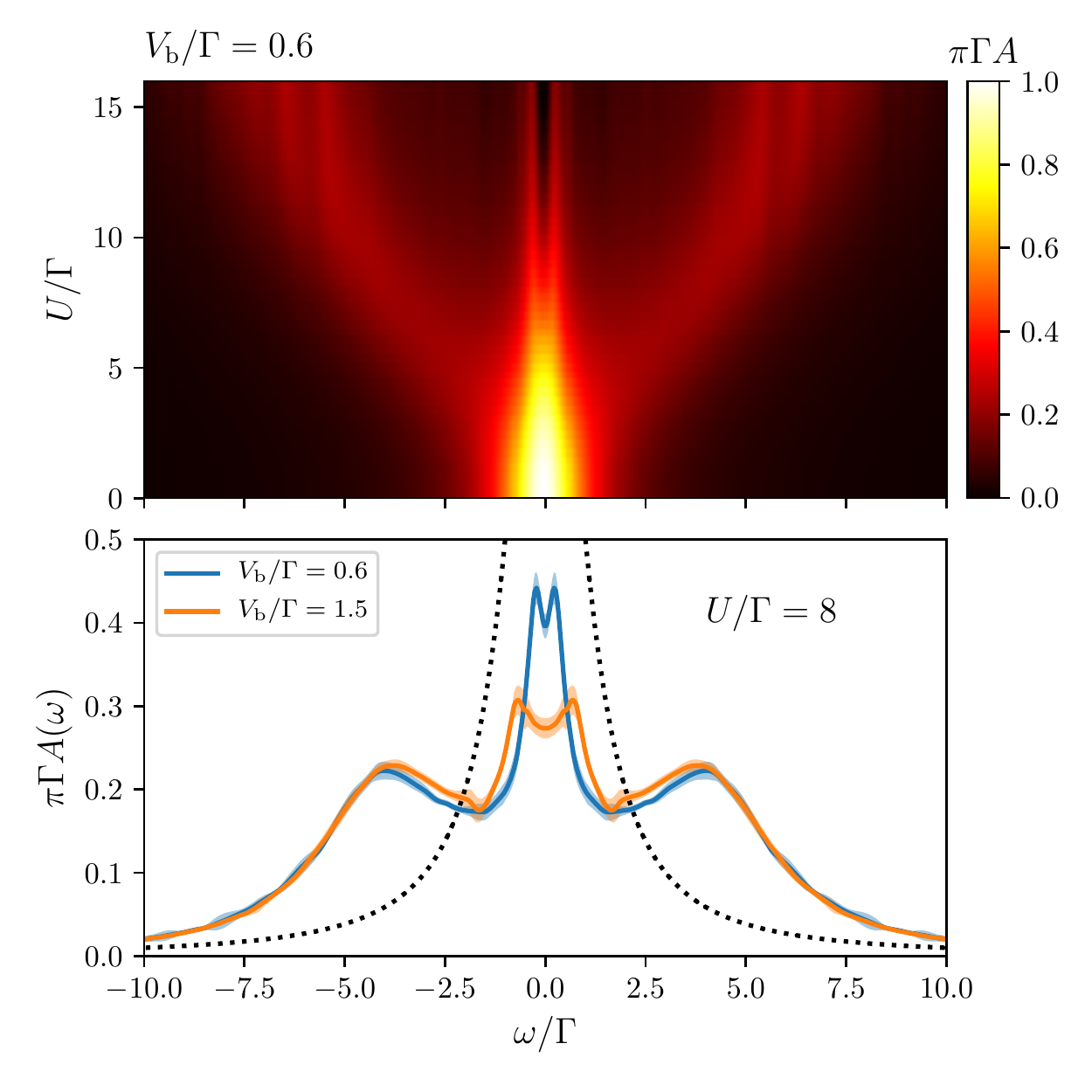}
    \caption{
    \label{fig:SF_bias_hot}
    Out-of-equilibrium spectral functions of the impurity with same parameters
    as in Fig.~\ref{fig:SF_sym_bias}, except for temperature $T = \Gamma / 50$.
    Upper panel: color plot of the spectral density as a function of $\omega$
    and $U$ for a voltage bias $V_{\rm b} = 0.6\Gamma$.  Lower panel: spectral
    density at $U=8\Gamma$ for a bias $V_{\rm b} = 0.6\Gamma$ (blue line) and
    $V_{\rm b} = 1.5\Gamma$ (orange line). Error bars are shown as shaded
    areas. The dotted line shows the non-interacting density.  No Bayesian
    inference has been used. Integration time is $20/\Gamma$.
    }
\end{figure}

Most of the results of this paper have been obtained at very low temperature.
We emphasize however that increasing the temperature makes the calculations easier:
indeed at finite temperature, the non-interacting Green's functions decrease exponentially 
as $e^{-t/T}$ instead of the algebric decay at zero temperature. It follows that the support of the integrals to be calculated is smaller, hence the convergence of the calculation faster.
We show a calculation at finite temperature in Fig.~\ref{fig:SF_bias_hot} where
we have computed the spectral density of the symmetric impurity at temperature $T = \Gamma / 50$
under a bias voltage $V_{\rm b} = 0.6\Gamma$ and $V_{\rm b} = 1.5\Gamma$. A
single Monte-Carlo run allows us to observe the splitting of
the Kondo resonance as $U$ is increased (upper panel). The result is
quantitatively accurate up to $U \approx 8\Gamma$ (lower panel) but remains
qualitatively meaningful at higher interaction (upper panel).

The fate of the Kondo resonance out-of-equilibrium,
in presence of a bias voltage, can be understood qualitatively from the
interplay of two phenomena. On the one hand, the bias voltage induces a
splitting of the Fermi energies of the two reservoirs, hence one expects a
corresponding splitting of the Kondo resonance. On the other hand, 
the voltage, like the temperature, increases the energy and phase space 
for the spin fluctuations, leading eventually to the disappearance of the
Kondo resonance\cite{Hershfield1991, Hershfield1992,Anders2008}. The
competition between both effects leads to the appearance of the splitting
only above a finite voltage threshold (about $V_b\simeq\Gamma$ in the plot 
of Fig.~\ref{fig:SF_sym_bias}).

\begin{figure}[htb]
    \centering
    \includegraphics[width=8cm]{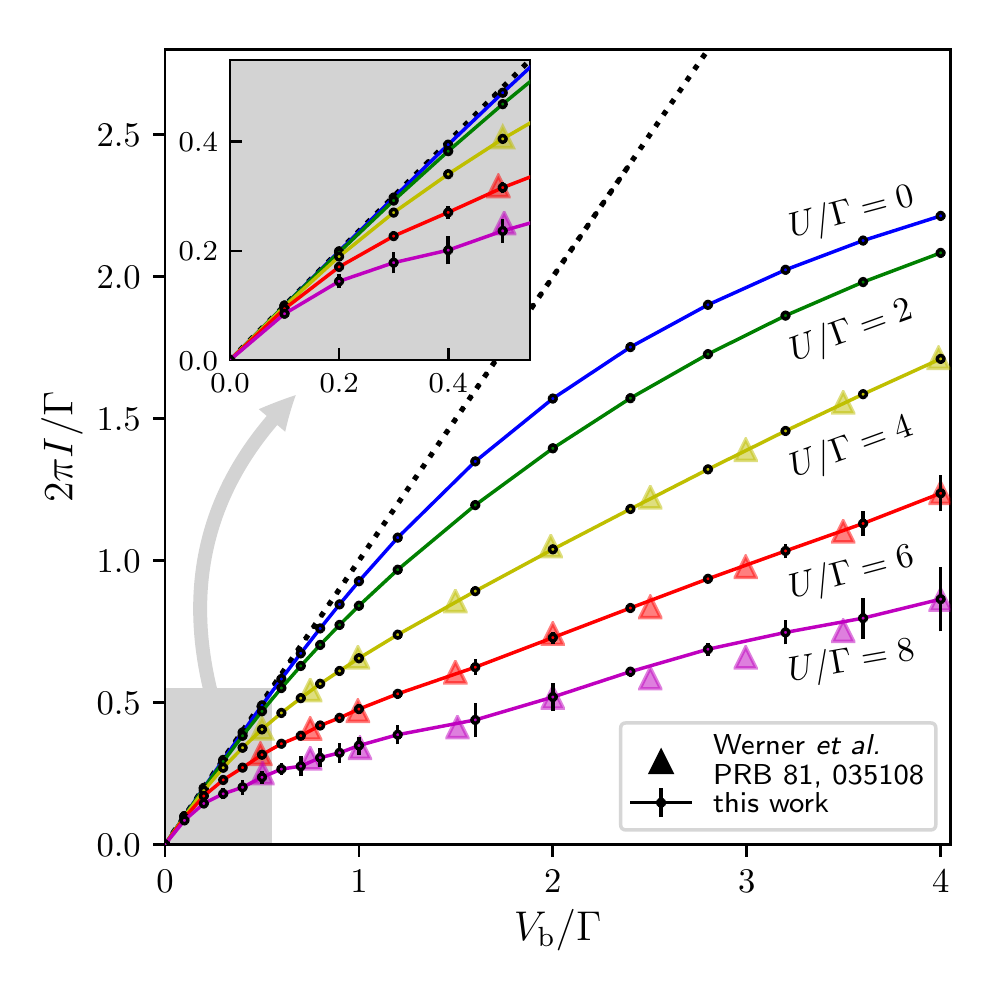}
    \caption{
    \label{fig:IV1}
    Current-voltage characteristics at different interaction strengths
    in the symmetric case $\epsilon_d=0$.
    Perturbation series for the current have been computed using the Landauer
    formula Eq.~(\ref{eq:landauer}), then resummed. The results are consistent with
    a weak-coupling Quantum Monte-Carlo calculation from
    Werner \textit{et al.}\cite{Werner2010} (triangles), but extends further down in bias.
    }
\end{figure}

\begin{figure}[htb]
    \centering
    \includegraphics[width=8cm]{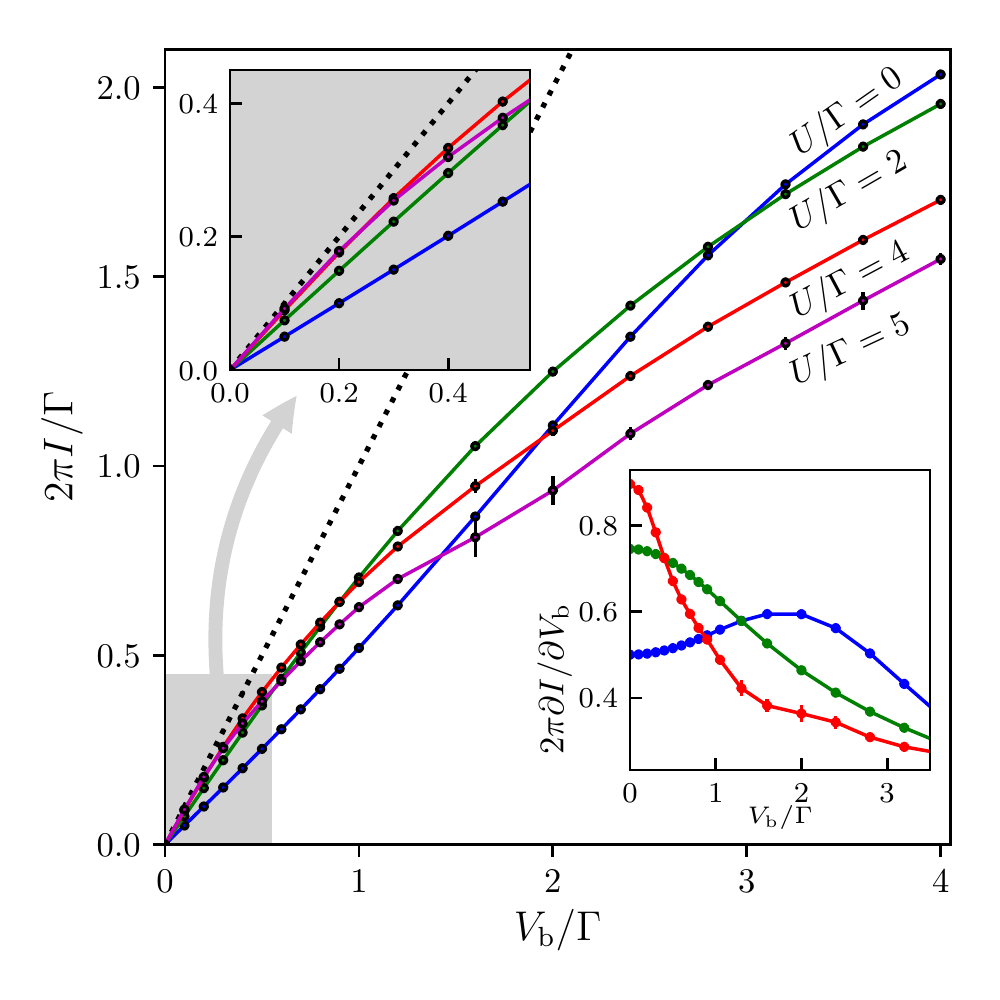}
    \caption{
    \label{fig:IV2}
    Current-voltage characteristics at different interaction strengths in 
    the asymmetric model ($\epsilon_d/\Gamma = 1$).
    The bottom-right inset shows the development of the zero-bias anomaly in
    the differential conductance when $U$ increases ($U/\Gamma = 0$ in blue,
    $2$ in green and $4$ in red).
    }
\end{figure}

\subsection{I-V transport characteristics}

Fig.~\ref{fig:IV1} shows the results obtained for the I-V characteristics in
the symmetric case $\epsilon_d=0$. The resummation has been done for the series of $1/I(U^2)$
using a parabolic transform with $p=-40\Gamma^2$. At small bias, we recover a
perfect transmission $I=(e^2/h) V_{\rm b}$ due to the unitary Kondo
resonance, while for $eV_{\rm b}> k_BT_K$ the conductance experiences an extra
suppression by the interaction (Coulomb blockade). We find a very good match with a previous calculation
from Ref.~\onlinecite{Werner2010}.  The present technique
allows one to lift the main limitations that Ref.~\onlinecite{Werner2009,Werner2010}
was facing: we can now access long times (here we have used $\sim 20/\Gamma$ but
it could be increased further if necessary) to be compared with maximum times
of the order of $\sim 3-5/\Gamma$ in Ref.~\onlinecite{Werner2010}. As a consequence,
we can reach the low bias regime, which was not accessible in
Ref.~\onlinecite{Werner2010}. Another important point is that the method is not
limited to the symmetry point as we now demonstrate.

Fig.~\ref{fig:IV2} shows the $I-V$ characteristics 
for an {\it asymmetric} model with $\epsilon_d/\Gamma=1$. 
The results have been obtained from the
resummation of $1/I(U)$ with a parabolic transform ($p=-6\Gamma$) and no
Bayesian inference. The $I-V$ characteristics is particularly interesting
because, due to the asymmetry, the non-interacting low bias transmission is
modified by interactions and one must first build up the Kondo resonance to approach 
$I\simeq (e^2/h) V_{\rm b}$
(note that the unitary limit is strictly exact only at $\epsilon_d=0$,
and the conductance is slightly lower than $e^2/h$ otherwise in the Kondo
regime).
This behavior leads to a non monotonous current versus $U$: as one
increases $U$, the current first increases until the Kondo resonance is fully built
 (see the bottom panel of Fig.~\ref{fig:SF_colormaps}).
As one increases further $U$, the Kondo width $T_K$ shrinks and
the current decreases as Coulomb blockade starts to set in.

\begin{figure}[htbp]
    \centering
    \includegraphics[width=8cm]{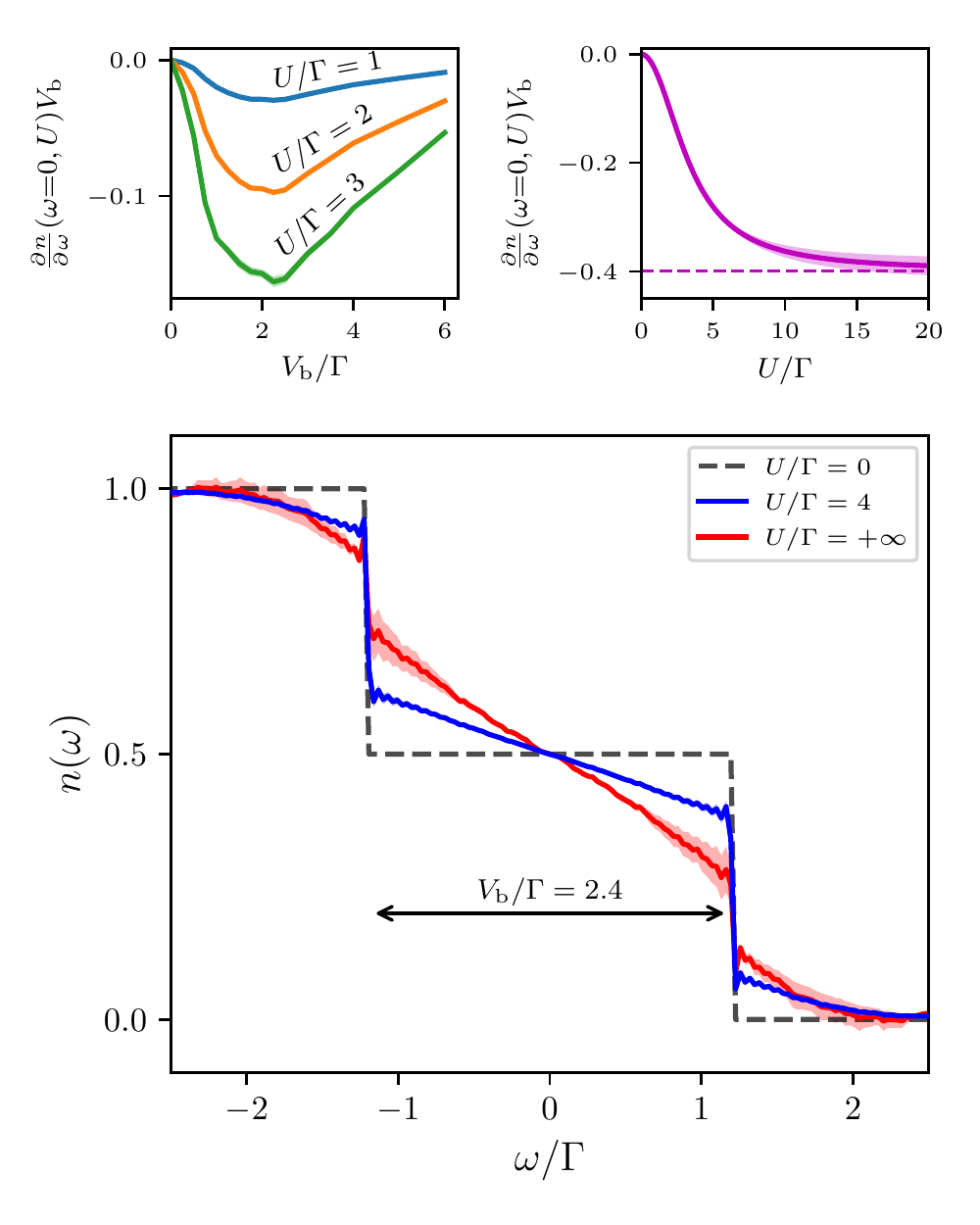}
    \caption{
    \label{fig:distribution_function}
    Lower panel: out-of-equilibrium electron distribution function on the
    impurity ($\epsilon_d/\Gamma = 0$) under a bias voltage $V_{\rm b} =
    2.4\Gamma$. The distribution function is defined as in
    Eq.~(\ref{eq:distrib_func_def}). Increasing the interaction strength
    ($U/\Gamma=4$ blue line, $U/\Gamma = +\infty$ red line) leads to a
    softening of the characteristic double-step of the non-interacting
    distribution function (dashed line). It is linear between the Fermi levels of the two leads. The Euler transform has been used for
    resummation and the result has not been submitted to Bayesian inference.
    Upper panels: normalized slope of the distribution function near $\omega=0$ as a function of bias voltage (left panel) and interaction strength (right panel).
    For intermediate interaction, the normalized slope reaches an extremum near $V_{\rm b} = 2 \Gamma$ (left panel).
    At strong interaction, the normalized slope saturates (right panel, for $V_{\rm b} = 2.4\Gamma$).
    }
\end{figure}

\subsection{Biased distribution function}

Finally, we discuss the out-of-equilibrium distribution function of the
impurity, \textit{i.e.} its energy-dependent probability of occupation.  We define the
distribution function $n(\omega)$ as 
\be
\label{eq:distrib_func_def}
n(\omega) = \frac{G^<(\omega)}{2\pi i A(\omega)},
\ee 
so that at equilibrium $n(\omega)$ is
simply the Fermi function $n_F(\omega)$. 
Without interaction, the distribution function amounts (at zero temperature) to a double step 
function
$n(\omega)_{U=0} = [n_F(\omega-V_{\rm b}/2)+n_F(\omega+V_{\rm b}/2)]/2$. 
We want to investigate the behaviour of $n(\omega)$ as $U$ increases, a
question that was not addressed in previous literature to the best of our
knowledge.

The results are shown in Fig.~\ref{fig:distribution_function}. In this
particular case, the series are fully alternated which means that the
singularity lies on the negative real axis. We could sum the series using an
Euler transform ($p=-8\Gamma^2$) up to $U=+\infty$. 
We find that the function $n(\omega)$ is not thermal, {\it i.e.} it can not be fitted 
by a Fermi function $n_F$ with an effective temperature.
In particular, it still exhibits discontinuities at the position of the lead Fermi surfaces, 
which we expect to be rounded at finite temperature. Interestingly, these
discontinuities are comparable to the equilibrium quasiparticle weight for
$U=4\Gamma$, do not seem to vanish in the limit $U=\infty$. Also very
striking is the quasi-linear behavior of $n(\omega)$ that is observed for 
$-V_b/2<\omega<V_b/2$.

Experiments that measure the non-equilibrium distribution function quantity typically use a 
third (for instance superconducting) terminal weakly coupled to the system 
\cite{Pothier1997, Anthore2003, Huard2004, Chen2009}. 
To the best of our knowledge, this quantity has not been measured in quantum
dots, and we hope that the present prediction may stimulate some experimental activity.


\section{Conclusion: The fall of the convergence wall}
\label{sec:conclusion}

We have presented a systematic computation of the perturbative expansion of the
Anderson impurity model in and out of equilibrium in power of the interaction
strength $U$. The main advantage of our Keldysh expansion approach is its
ability to calculate directly in the long time steady state regime.
Using our approach, we were able to obtain improved or novel results regarding the
non-equilibrium dynamics of strongly interacting quantum dots.

The main contribution of this article lies in the systematic construction of a
set of conformal transformations that provide a practical route for a
mathematically controlled resummation of series.  We have shown how to use
analytic conformal transform guided by an approximated location of the
singularities of the physical quantities in the $U$ complex plane.  We also
presented a Bayesian method to control the strong amplification of statistical
noise during this procedure, using some simple non-perturbative information.
The combination of singularity location, conformal transform crafting and
Bayesian inference provides a robust and generic resummation methodology. 

It was noticed recently\cite{Rossi2017a} that for values of $U$ {\it inside}
the convergence radius, connected diagrammatic quantum Monte-Carlo techniques
provide a systematic route for calculating the many-body quantum problem in a
computational time that only increases {\it polynomially } with the requested
precision. We argue that the argument of Ref.~\onlinecite{Rossi2017a} can be directly
extended to systems where the separation hypothesis holds (switching from
working with the series in $U$ to the series in $W$). We conclude that, in
general, systems where the separation hypothesis hold can be computed with a
computing time that increases polynomially with the requested precision.

The approach presented here may have implications for a large class of other
problems within or beyond condensed matter physics. In particular, a possible
extension is to build a real time (equilibrium or non-equilibrium) quantum
impurity solver for DMFT or its extensions, or directly addressing lattice
problems such as the Hubbard model.  At its core, it consists in techniques to
efficiently compute the bare perturbation series and to sum it. 
Its limitations remain to be explored.  They could come from a resurgence of the sign problem,
which would manifest itself in a very oscillatory nature of the integrals for
expansion coefficients, making them hard to evaluate, or from a difficulty to
sum the perturbative series, in particular for systems with a phase transition,
or a non-Fermi liquid fixed point at low temperature.
In order to address these questions, the technique needs to be applied
to more complex models. Work is in progress in this direction.

\begin{acknowledgments}
The Flatiron Institute is a division of the Simons Foundation.
We acknowledge useful discussions with Laura Messio, Volker Meden and Christophe Mora.
We thank our anonymous referee for pointing out Ref.\cite{Horvatic1985}.
We acknowledge financial support from the graphene Flagship (ANR FLagera GRANSPORT),
the French-US ANR PYRE and the French-Japan ANR QCONTROL.
\end{acknowledgments}


\begin{figure}[ht]
    \centering
    \includegraphics[width=8cm]{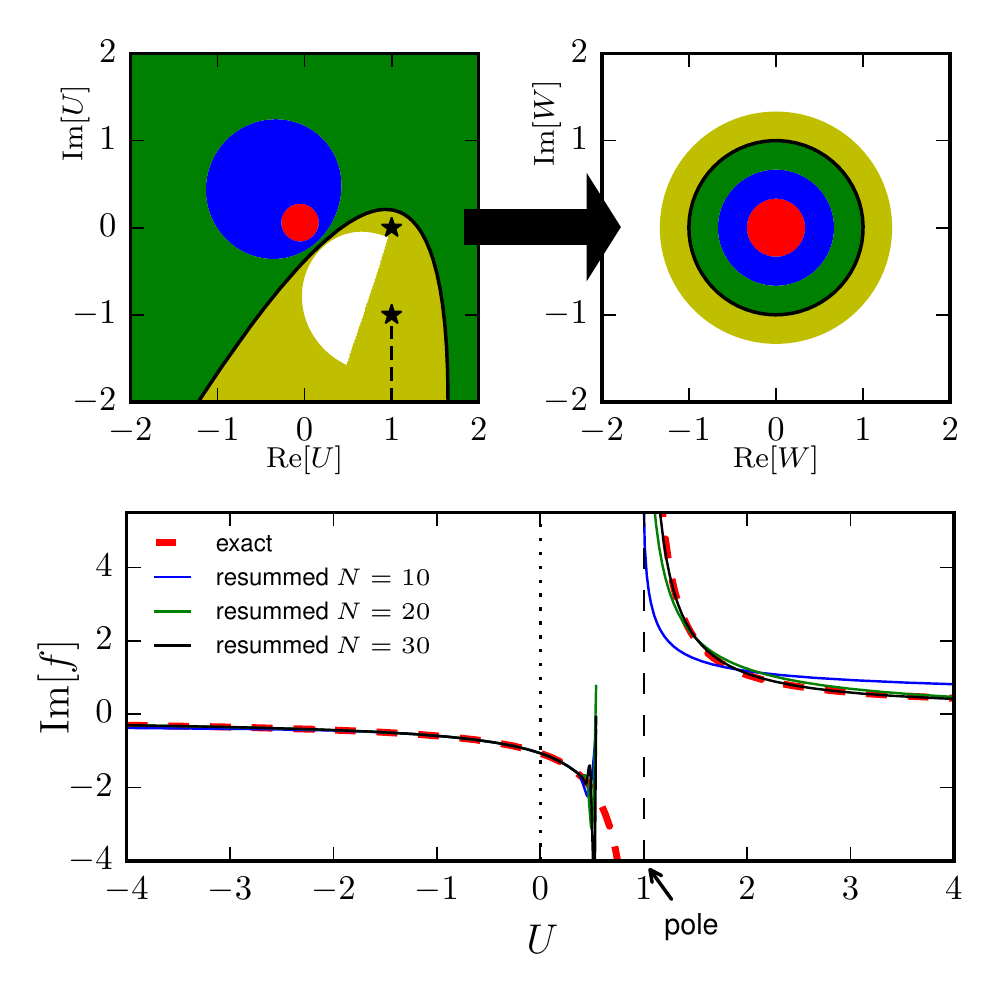}
    \caption{
        \label{fig:resum_past_pole}
        Resummation of the series of $f(U) = 1/\ln(i(1-U)+1)$ near $U=0$ on the real positive axis, beyond the pole at $U=1$.
        $f$ has a pole at $U = 1$ and a branch cut starting at $U=1-i$ and going in straight line toward $1-i\infty$ (stars and dashed line in upper left panel).
        We isolated these singularities by applying a conformal map $W = \frac{\chi(U) - \chi(0)}{\chi(U) - \chi(0)^*}$, with $\chi(U) = i\sqrt{(U-1)/p} - i$.
        It maps the inside of a parabola into the outside of the unit disk (left and right upper panels).
        $p$ controls the direction and width of the parabola.
        Here $p = 0.2 e^{i\pi\times 0.4}$.
        With $N=30$ terms, one can compute $f$ for all real $U$ (black plain line) except a narrow band around the pole (dashed black vertical line).
    }
\end{figure}
\appendix
\section{A toy model function with a singularity on the real axis}
\label{app:beyond_singularity}

We present here on  Fig.~\ref{fig:resum_past_pole} 
a toy model for the resummation of a function 
$f(U)= 1/\ln(i(1-U)+1)$ that has a pole on the real axis at $U=1$ as well as a branch
cut on the curve $U=1-i (1+x)$ with $x\in [0,\infty]$. The aim of this toy
model is to show that even though $f(U)$ has a singularity on the real axis
(and hence it will be difficult to calculate close to this singularity), it is
possible to calculate the function beyond the singularity using a conformal
transformation. We use the conformal map $W = \frac{\chi(U) - \chi(0)}{\chi(U)
- \chi(0)^*}$, with $\chi(U) = i\sqrt{(U-1)/p} - i$ that maps the inside of a
parabola into the outside of the unit disk (see the upper left and right panels
of Fig.~\ref{fig:resum_past_pole}).  The lower panel of
Fig.~\ref{fig:resum_past_pole} shows the corresponding resummed series using
$N=10, 20$ and $30$ terms in the expansion of $f(U)$.  Although we cannot
calculate close to $U=1$, we find that with as little as $N=20$ terms in the
expansion of $f(U)$, we can recover an accurate description of $f(U)$ for
$U>1.2$ from an expansion around $U=0$.


\section{Convergence of the perturbation series at finite time}
\label{app:convergence_finite_time}

In this appendix, we show that {\it at finite} time $t$, the radius of convergence of the perturbation series for an operator $\cal{O}$ is infinite, 
for a system with an interaction on a  finite number of sites and an infinite bath. 
Indeed, the average is given by 
\begin{equation}\label{app2.exp1}
\langle{\cal \hat O}(t) \rangle \propto \left\langle T_c e^{ -i U\int du \hnd{H}_{\rm int}(u)} 
{\cal \hat O} (t)
\right\rangle,
\end{equation}
where the integral goes along the forward-backward Keldysh contour $0 \rightarrow t \rightarrow 0$, 
the operators are taken in the interaction representation, $ T_c$ is the
usual Keldysh contour ordering operator and $\hnd{H}_{\rm int}(u)$ is the
interacting part of the Hamiltonian.

More precisely, each of the $2^n$ terms of the expansion of the exponential has the form,
\begin{equation*}
\frac{U^n}{n!} \int_{[0,t]^n} du_1...du_n \langle {\cal \hat O} (t) C(u_1,\ldots u_n)\rangle.
\end{equation*}
where $C$ is a product of $c$, $c^\dagger$, and unitary time evolution operators.
The terms $\langle {\cal \hat O} (t) C(u_1,\ldots u_n)\rangle$ are amplitudes of probability for quantum processes and are therefore bounded.
Explicitly, 
\begin{align*}
    \langle {\cal \hat O} (t) C(u_1,\ldots u_n)\rangle &=
   \mathrm{Tr} \left( \frac{e^{-\beta H_0}}{Z_0}  {\cal \hat O} (t) C(u_1,\ldots u_n) \right) \\
   &=  \sum_\psi \langle \psi | \frac{e^{-\beta H_0}}{Z_0} {\cal \hat O} (t) C(u_1,\ldots u_n) | \psi \rangle 
\end{align*}
\begin{align*}
   \left | \langle {\cal \hat O} (t) C(u_1,\ldots u_n)\rangle \right |
   &\leq
   \sum_\psi  \frac{e^{-\beta H_0}}{Z_0} \norm{ {\cal \hat O} (t) C(u_1,\ldots u_n)  \psi } 
   \\
   &\leq  
   \sum_\psi  \frac{e^{-\beta H_0}}{Z_0} \norm{ {\cal \hat O} (t) C(u_1,\ldots u_n)}  \norm{\psi }
   \\
   &\leq \norm{ {\cal \hat O} },
\end{align*}
where $\norm{v}$ is the norm for a vector and the induced norm for an operator.
We note that the norm is not modified by the unitary evolution $\norm{e^{iH_0 u} Ae^{-iH_0 u} } = \norm{A}$ for any operator $A$, 
and for the canonical operators $\norm{c} = 1$, as can be checked in the Fock basis, {\it independently of the size of the bath}.
Since the norm is sub-multiplicative, we obtain the last inequality.

Therefore, the term of order $n$ in the expansion of (\ref{app2.exp1}) is controlled by 
a bound $\frac{\norm{{\cal O}}(2 U t L)^n}{n!}$, so the series
has an infinite radius of convergence. Note that this argument is valid because
the electron-electron interaction is present on a finite number of sites only.
It would not apply directly to e.g. the Hubbard model in the thermodynamic limit.

\bibliographystyle{apsrev}
\bibliography{../diag_MC_Anderson}{}

\end{document}